\documentclass[aps,showpacs,twocolumn,longbibliography,nofootinbib,secnumarabic]{revtex4-1}
\usepackage{graphicx}% Include figure files
\usepackage{dcolumn}% Align table columns on decimal point
\usepackage{bm}% bold math
\usepackage{braket}
\usepackage{leftidx}
\usepackage{cancel}
\usepackage{amsmath}
\usepackage{epstopdf}
\usepackage{color}
\usepackage[mathcal]{eucal}
\usepackage{float}

\usepackage{yfonts}
\usepackage{color}

\usepackage{scalerel}

\begin{document}
\date{\today}
%Title of paper
\title{Negative superluminal velocity  and violation of Kramers-Kronig relations \\ in ``causal" optical setups}

\author{Mehmet Emre Tasgin}%\thanks{metasgin@hacettepe.edu.tr }

\affiliation{Institute of Nuclear Sciences, Hacettepe University, 06800, Ankara, Turkey}
\affiliation{metasgin@hacettepe.edu.tr and metasgin@gmail.com}

%nonanalyticity
%nonanalyticities

\begin{abstract}
We investigate nonanalyticities (e.g., zeros and poles) of refractive index~$n(\omega)$ and group index~$n_g(\omega)$ in different optical setups. We first demonstrate that: while a Lorentzian dielectric has no nonanalyticity in the upper half of the complex frequency plane~(CFP), its group index ---which governs the pulse-center propagation--- violates the Kramers-Kronig relations~(KKRs). Thus, we classify the nonanalyticities as in the (a) first-order (refractive index or reflection) and (b) second-order (group index or group delay). The latter contains the derivative of the former. Then, we study a possible connection between the negative superluminal velocities and the presence of nonanalyticities in the upper half of the CFP. We show that presence of nonanalyticities in the upper half of the CFP for (a) the first-order response and (b) the second-order response are accompanied by the appearance of negative (a) phase velocity and (b) group velocity, respectively. We also distinguish between two kinds of superluminosity, $v>c$ and $v<0$, where we show that the second one~($v<0$) appears with the violation of KKRs.
\end{abstract}

\maketitle

\section{Introduction} \label{sec:intro}

Possibility of superluminal propagation is an old, but still an open, question on which intriguing research continues~\cite{ChuPRL1982SL,WangNature200SL,chiao1999tunneling,talukder2005measurement,peatross2000average,nanda2009superluminal,talukder2014direct,boyd2009controlling}. Tracking the pulse peak, Poynting vector or energy average, all, demonstrate superluminal propagation~\cite{peatross2000average,nanda2009superluminal,talukder2014direct,kohmoto2005nonadherence}, thus confirm the observation of superluminosity in experiments~\cite{ChuPRL1982SL,WangNature200SL,chiao1999tunneling,talukder2005measurement}. These pulse-center tracking methods for defining the propagation velocity, however, are shown as not-reliable measures for a propagation speed~\cite{talukder2014direct,tasginPRA2012testing}. If they would truly correspond to a realistic flow of the information, the response function should present nonanalyticities in the upper half~(UH) of the complex frequency plane~(CFP), UH-CFP~\footnote{That is, response function should not vanish for advanced times~\cite{Jackson_book}, i.e., by displaying nonanalyticities in the UH-CFP. \label{fn:Gtau}}. Indeed, some recent studies~\cite{mandelstam1971lectures,stenner2003speed,shore2007superluminality,zhu2003propagation} show that the wave-front~(signal) velocity, an infinitesimally sudden disturbance, does not exceed $c$, the vacuum velocity of light, in these media. 

Apart from the pulse-center superluminosity, there exist optical setups whose transfer functions (e.g., reflection) display nonanalyticities in the UH-CFP~\cite{wang2002causal,beck1991group,stern2012transmission,wangPRA2007theoretical}. In such ``causal" devices~\cite{wang2002causal,beck1991group,stern2012transmission,wangPRA2007theoretical}, like a Gires-Tournois interferometer~\cite{beck1991group} and an all-pass filter~\cite{wang2002causal,stern2012transmission}, multiple interference effects can make the transfer functions violate the Kramers-Kronig relations~(KKRs) by introducing nonanalyticities in he UH-CFP. Actually, this should not be too surprising. Because, we already know that, amplitude of a wave (radiation or particle) can vanish in a spatial region via interference at specific wavelengths. Such a phenomenon, most possibly occurring owing to the (assumption of) instantaneous spreading of the wavefunction~\footnote{That is, in the wave mechanics treatment we already assume such interferences happen before the propagation of the pulse~\cite{dressel2008kramers}.}
 to infinity~\cite{hegerfeldt1998instantaneous,hegerfeldtPRD1974remark}, appears also in relativistic equations~\cite{perezPRD1977localization,hegerfeldtPRD1980remarks}.

In such setups, defining a tunneling time can also be problematic~\cite{davies2005quantum,gasparian1998application,gruner1997photon} since wavepacket appears to tunnel barriers of different thicknesses at equal time~\cite{hartman1962tunneling}. Tough relativistic field theory treatments of evanescent waves exhibit superluminal transport~\cite{winfulPRL2003,winfulNature2003}, without violating the weak causality~\footnote{Weak causality is the satisfaction of Einstein causality for expectation values or ensemble averages merely, but not for each individual process~\cite{hegerfeldt1998instantaneous}.}, again a superluminal propagation is discussed to be not possible in such systems~\cite{winfulPRL2003,winfulNature2003}.

Recent studies show that violation of Kramers-Kronig relations~(KKRs), i.e., nonclassicalities being located in the UH-CFP, can appear also in an Otto configuration~\cite{wang2016counterintuitive} and in a gain slab~\cite{Zubairy2014counterintuitive}~\footnote{\label{fn:newKKR} Although a new method~\cite{kop1997kramers} is derived for relating real and imaginary parts of the refractive index, for application (experimental) purposes; violation of KK relations, i.e. $G(\tau)\neq 0$ for $\tau<0$, is not circumvented yet. M.Suhail Zubairy--- private communication.}. In addition to such interference-origined setups~\cite{wang2002causal,beck1991group,stern2012transmission,wangPRA2007theoretical,wang2016counterintuitive,Zubairy2014counterintuitive}, in this work we explore that transfer functions of an optomechanical system can also violate the KKRs. We show that in an optomechanical system, zeros of reflection/transmission $R(\omega)$,$T(\omega)$ move to the UH-CFP by increasing the cavity-mirror coupling over a critical value $g>g_{\rm crt}$.

The studies, we mention above, explore the nonanalyticities of a single wavelength (e.g., phase velocity) response. The superluminosity, observed in the experiments~\cite{ChuPRL1982SL,WangNature200SL,chiao1999tunneling,talukder2005measurement} and theories~\cite{peatross2000average,nanda2009superluminal,talukder2014direct,boyd2009controlling}, however, demonstrate superluminal pulse-center propagation which is governed by the group behavior. Group velocity contains the derivative of the refractive index, i.e., $v_g=d\omega/dk$ and $n_g(\omega)=n(\omega)+\omega dn/d\omega$~\cite{kohmoto2005nonadherence,Jackson_book}. Thus, if one aims to explore the nonanalyticities corresponding to the superluminal propagation observed in Refs.~\cite{ChuPRL1982SL,WangNature200SL,chiao1999tunneling,talukder2005measurement,peatross2000average,nanda2009superluminal,talukder2014direct,boyd2009controlling}, she/he needs to investigate the nonanalyticities of $n_g(\omega)$.

In this paper, we investigate the nonanalyticities (zeros/poles) of both (i) first-order response~(can be refractive index or reflection/transmission functions) and (ii) second-order response~[can be group index or group delay $\tau_R(\omega)$]~\footnote{\label{fn:name_1st_2nd_order} We name the single frequency (refractive index, phase-velocity) response as the first-order and the group (wave packet) behavior as the second-order since the latter contains derivative of the former one. We also aim to introduce a notion for the classification of the two situations.}. We study (a) a uniform Lorentzian dielectric, (b) reflection/transmission from/by (b) an absorbing slab, (c) a periodic structure, (d) Otto configuration, and (e) and optomechanical system~\footnote{Nonanalyticities of an optomechanical system are studied here for the first time, to our best knowledge.}. We also present a differentiation between two kinds of superluminousity, $v_g>c$ and $v_g<0$ (or $\tau_R<0$).

First, we consider a uniform Lorentzian dielectric medium. We notice that when the nonanalyticities of the group index $n_g(\omega)$ move into the upper half of the CFP~(UH-CFP), the group velocity becomes negative $v_g<0$. We also observe that for $v_g>c$, nonanalyticieties of $n_g(\omega)$ remain in the lower half of the CFP. Thus, we decide that positive superluminosity (i.e. $v_g>c$) in a Lorentzian medium~\cite{ChuPRL1982SL,WangNature200SL,chiao1999tunneling,talukder2005measurement,peatross2000average,nanda2009superluminal,talukder2014direct,boyd2009controlling} can be deemed analogous to $v=c/n(\omega)>c$ in a first-order (single frequency wave) response.

The observation ``$v_g<0$ in a Lorentzian medium when nonanalyticities of $n_g(\omega)$ move into the UH-CFP" triggers us for further research. Systems (b)-(e) are also known for exhibiting {\it negative group} delays in the reflected/transmitted pulses~\cite{wang2006superluminal,hwang2006correlation,wang2016counterintuitive,tarhan2013superluminal,wang2017tunable}. So, we also investigate the nonanalyticities of the second-order response ($\tau_R(\omega)=d\phi_R/d\omega$ or $n_g^{(\rm eff)}=n_{\rm eff} + \omega \: dn_{\rm eff}/d\omega$) in the setups (b)-(e). Here, $\phi_{R,T}$ is the phase of the reflected/transmitted wave and $n_{\rm eff}(\omega)$ is an effective refractive index~\cite{smith2002determination,yoo2019causal}. We find that, also in systems (b)-(e), where $\tau_R<0$, nonanalyticities of the second-order (group) response lye in the UH-CFP.

This result further intrigues us if the situation is similar for the first-order (single frequency wave) response, e.g., in $R(\omega)$ or $T(\omega)$.  We find that an abrupt sign change in the phase-velocity (single frequency phase delay $\phi_R$) and the effective index $n_{\rm eff}(\omega)$, from positive to negative, accompanies the movement of the nonanalyticities of the first-order response to the UH-CFP. This happens both in (d) Otto configuration and (e) optomechanical setup. We remark that the systems (a)-(c) already do not possess any nonanalyticity in the UH-CFP regarding the first-order response. 

This interesting result is also partly connected with the search for negative index materials other than both $\mu<0$,$\epsilon<0$~\cite{mackay2009negative,depine2004new,kinsler2008causality,stockman2007criterion}~\footnote{\label{fn:negative-index mater} We remark that these references discuss the case where group and phase velocities are counter directed, i.e., $v \times v_g<0$, where as in our results both $v<0$ and $v_g<0$.} and with the relationship between negative index and KKRs~\cite{nazarov2015negative,akyurtlu2010relationship,dolgov1981admissible,klimchitskaya2018kramers}. Direct observation of the negative phase velocity in 2D hexagonal boron nitride~(h-BN), via ultrashort scanning near field optical microscopy~(SNOM) techniques~\cite{yoxall2015direct}, promotes new research in this field. 

It is also worth noting that ``none" of the setups (a)-(d), we study here, exhibit gain which could be argued as being responsible for nonanalyticities in the UH-CFP. While (e) an optomechanical system~\cite{tarhan2013superluminal,agarwal2010electromagnetically,genes2008robust,vitaliPRL2007optomechanical} makes use of a coupler (pump) laser for tuning the effective cavity-mirror interaction $g$, we demonstrate that the form of this effective interaction hamiltonian (actually) does not necessitate the presence of gain.

The paper is organized as follows. In Sec.~\ref{sec:phase_velocity_etc}, we revise the insight for phase and group velocity, and Kramers-Kronig relations. In Sec.~\ref{sec:effective_index}, we introduce the effective index method which we use as an alternative approach for demonstrating negative velocities. We first study the second-order response, Sec.~\ref{sec:second-order}. We  study the setups (a) uniform Lorentzian dielectric, (b) absorbing slab and (c) reflection form a periodic structure which are known to exhibit negative group delay $\tau_R<0$. We find that the negative group delays~\cite{wang2006superluminal,hwang2006correlation,wang2016counterintuitive,tarhan2013superluminal,wang2017tunable}, actually, accompany the presence of nonanalyticities in the UH-CFP. In Sec.~\ref{sec:first-order}, we investigate if such a situation exists also in the first-order response. We show that, indeed, a negative phase-velocity accompanies the movement of nonanalyticities to the UH-CFP both in (d) Otto configuration and (e) optomechanics. Sec.~\ref{sec:conclusion} contains our summary and conclusions.

\section{A Short Review of Background Information} \label{sec:fundamental}

%%%%%%%%%%%%%%%%%%%%%%%%%%%%%%%%%%%%%%%%%%%%%%%%%%%%%%%%%%%%%%%%%%%%%%%%%%%%%%%%%%%%%%%%%%%%%%%%%%%%%%%%%%%%%%%%%%%%%%%%%%%%%%%%%%%%%%%%%%%%%%%%%%%%
\subsection{Phase velocity, Group velocity and Kramers-Kronig relations} \label{sec:phase_velocity_etc}

In this section, we revise the notion of group velocity and group delay $\tau_{R,T}(\omega)$ via a widely preferred illustrative approach. We pay attention to $\tau_{R,T}(\omega)$, because we determine the nonanalyticities of $n_g(\omega)$ and $\tau_{R,T}(\omega)$ in the following sections.

Phase velocity of a single frequency sine wave $\sin(kx-\omega t)$ is set by the refractive index of the medium, i.e., $x/t=\omega/k=c/n(\omega)$. When several numbers of such sine waves superpose (a wave packet)
\begin{eqnarray}
E(x,t)=E_0 \left[ \sin(k_1 x-\omega_1 t) + \sin(k_1 x-\omega_1 t) \right] \nonumber
\\
=E_0\sin(kx-\omega t) \: \cos(\Delta k \: x - \Delta \omega \: t), \label{vg_envelop}
\end{eqnarray}
two velocities introduce. $v=\omega/k$ is the phase velocity while $v_g=\Delta\omega/\Delta k$ is the envelope (group) velocity. Here, $\Delta k=(k_2-k_1)/2$, $\Delta \omega=(\omega_2-\omega_1)/2$ and $k=(k_2+k_1)/2 \simeq k_{1,2}$, $\omega=(\omega_2+\omega_1)/2\simeq \omega_{1,2}$. The second velocity $v_g$ determines the propagation of the envelop or the pulse-center~\cite{peatross2000average,nanda2009superluminal,talukder2014direct,boyd2009controlling,kohmoto2005nonadherence}. Tough behavior of the wave packet propagation can also be demonstrated via rigorous derivations, e.g., in common textbooks~\cite{Jackson_book}, here we rather prefer the illustrative approach given in Eq.~(\ref{vg_envelop}).

Similar to the refractive index $n(\omega)=k/\omega$, the group index $n_g(\omega)=dk/d\omega$ needs not possess any nonanalyticity in the UH-CFP. Otherwise, an event which has not happened yet would affect the pulse-center~(second-order) propagation, i.e., $G(t-t')\neq 0$ for $t'>t$~\cite{Jackson_book}. As the Kramers-Kronig relations~(KKRs) are obtained assuming that all the nonclassicalities lye in the lower half of the CFP, pulse-center propagation would also violate the KKRs. Although KKRs can be ``reorganized" for obtaining the real part of the index from the absorption (imaginary part)~\cite{kop1997kramers}~${}^{\ref{fn:newKKR}}$, the problem with the causal structure of the response function remains~\footnote{\label{fn:causal} We kindly underline that: by calling ``violation of KKRs" we do not imply the ``violation of causality" in such devices. Calling as ``violation of KKRs" we always have in mind the existence of a method/reason (yet not known) to circumvent from ``violation of KKRs" implying ``violation of causality". This, for instance, can be a flaw due to the instantaneous spreading of the wavefunctions to infinity~\cite{hegerfeldt1998instantaneous,hegerfeldtPRD1974remark}. These devices~\cite{wang2002causal,beck1991group,stern2012transmission,wangPRA2007theoretical,wang2016counterintuitive} are already ``causal" devices. As a further example, in the optomechanical setup, violation of KKRs can be circumvented~\cite{tasgin2019entanglement} by admitting the results of Refs.~\cite{sonner2013holographic,jensen2013holographic,Susskind2013cool} and \cite{hollowood2008refractive,hollowood2007causality}.  }.

When the medium is not a uniform one, but contains reflecting/transmitting layers, e.g., an absorbing slab in Fig.~\ref{fig3}; the group delay $\tau_R$ can be determined from the phase of the reflected wave~\cite{wang2006superluminal,agarwal2010electromagnetically,tarhan2013superluminal} $R(\omega)=|R(\omega)|e^{i\phi_{\scriptscriptstyle R}}$ as follows. Again, superposition of two close-lying frequencies at a fixed position $x=0$ can be written as
\begin{eqnarray}
E(t)=E_0 \sin(-\omega_1 t+ \phi_1) + E_0 \sin(-\omega_2 t+ \phi_2) \nonumber
\\
=E_0 \sin(-\omega t + \phi) \: \cos(\Delta \omega \, t - \Delta\phi), \label{tauR_envelop}
\end{eqnarray}
where the movement of the envelop (group propagation) is determined again by the cosine part~\cite{wang2006superluminal,agarwal2010electromagnetically,tarhan2013superluminal}. Hence the group delay around frequency $\omega$ is $\tau_{\scriptscriptstyle R}(\omega)=\left( d\phi_{\scriptscriptstyle R}/d\omega' \right)_\omega$. A rigorous derivation can also be performed via the same treatment given in standard textbooks~\cite{Jackson_book}.  Similar to $n_g(\omega)$ in a uniform medium, $\tau_{R}(\omega)$ needs to satisfy the KKRs in a setup divided into different indexed materials. 

We investigate the first and second-order nonanalyticities of the systems (b)-(e) by exploring the zeros/poles of $R(\omega)$,$T(\omega)$) and $\tau_R(\omega)$, respectively.

%%%%%%%%%%%%%%%%%%%%%%%%%%%%%%%%%%%%%%%%%%%%%%%%%%%%%%%%%%%%%%%%%%%%%%%%%%%%%%%%%%%%%%%%%%%%%%%%%%%%%%%%%%%%%%%%%%%%%%%%%%%%%%%%%%%%%%%%%%%%%%%%%%%%
\subsection{Effective index Method} \label{sec:effective_index}

We also use an effective index $n_{\rm eff}(\omega)$ method in exploring the nonanalyticities of the setups (b)-(e), both for phase and group response, in addition to $R(\omega)$ and $\tau_R(\omega)$. We adapt the effective index as an alternative method to double-check the presence of nonanalyticities in the UH-CFP and the negative velocities. Effective index method, we summarize below, is commonly utilized for various optical elements, especially for metamaterials~\cite{smith2002determination,yoo2019causal}.

The method is quite straightforward. There is an optical element whose reflection $R(\omega)$ and/or transmission $T(\omega)$ coefficients are known. The question is: what kind of a refractive index $n_{\rm eff}(\omega)$ one can assign to this optical element such that the index $n_{\rm eff}(\omega)$ results the same $R(\omega)$ and $T(\omega)$ coefficients.  We note that this is a strong restriction on $n_{\rm eff}(\omega)$ since it has to match the optical element for all $\omega$ values. The details of the method can be found in Refs.~\cite{smith2002determination,yoo2019causal}.

Assuming a nonmagnetic medium, which is the case in most optical elements, the effective dielectric function can be determined~\cite{smith2002determination,yoo2019causal} as 
\begin{equation}
\epsilon_{\rm eff}(\omega) = \frac{(R(\omega)-1)^2 + \tilde{T}^2(\omega)}{(R(\omega)+1)^2 - \tilde{T}^2(\omega)} 
\label{effective_eps}
\end{equation}
for a finite thickness $L$ optical element with $\tilde{T}(\omega)=T(\omega) e^{ikL}$.

When the optical element occupies a semi-infinite space, e.g., reflection from a semi-infinite periodic structure in Fig.~\ref{fig5} or the Otto configuration in Fig.~\ref{fig7}, effective dielectric function can be found simply by equating the reflection coefficient $R(\omega)$ to the reflection formula $r(\omega)$ given in standard textbooks~\cite{Jackson_book,griffiths_book} for various incidence conditions. In obtaining the effective group index $n_g^{\rm (eff)}(\omega)$, we simply use $n_g^{\rm (eff)}(\omega)=n_{\rm eff}(\omega) + \omega\: dn_{\rm eff}/d\omega$.

%%%%%%%%%%%%%%%%%%%%%%%%%%%%%%%%%%%%%%%%%%%%%%%%%%%%%%%%%%%%%%%%%%%%%%%%%%%%%%%%%%%%%%%%%%%%%%%%%%%%%%%%%%%%%%%%%%%%%%%%%%%%%%%%%%%%%%%%%%%%%%%%%%%%
%%%%%%%%%%%%%%%%%%%%%%%%%%%%%%%%%%%%%%%%%%%%%%%%%%%%%%%%%%%%%%%%%%%%%%%%%%%%%%%%%%%%%%%%%%%%%%%%%%%%%%%%%%%%%%%%%%%%%%%%%%%%%%%%%%%%%%%%%%%%%%%%%%%%
%%%%%%%%%%%%%%%%%%%%%%%%%%%%%%%%%%%%%%%%%%%%%%%%%%%%%%%%%%%%%%%%%%%%%%%%%%%%%%%%%%%%%%%%%%%%%%%%%%%%%%%%%%%%%%%%%%%%%%%%%%%%%%%%%%%%%%%%%%%%%%%%%%%%
\section{Nonanalyticities in the Second-Order response} \label{sec:second-order}

In this section, we investigate nonanalyticities of the second-order (group) response in (a) a uniform Lorentzian dielectric medium, reflection/transmission by/through (b)  an absorbing slab and (c) a semi-infinite periodic structure. We demonstrate that the observed negative velocities~\cite{wang2006superluminal,hwang2006correlation} actually accompany the presence of nonanalyticities in the UH-CFP for the second-order (group, wave packet) response functions.

%%%%%%%%%%%%%%%%%%%%%%%%%%%%%%%%%%%%%%%%%%%%%%%%%%%%%%%%%%%%%%%%%%%%%%%%%%%%%%%%%%%%%%%%%%%%%%%%%%%%%%%%%%%%%%%%%%%%%%%%%%%%%%%%%%%%%%%%%%%%%%%%%%%%
\subsection{Uniform Lorentzian dielectric}  \label{sec:Lorentzian}

As the first example, which triggered the research in this work, we consider (a) a uniform medium filled with a Lorentzian dielectric function~\cite{tanaka1986propagation}
\begin{equation}
\epsilon(\omega) = \epsilon_{c} + f\frac{\omega_0^2}{\omega_0^2-\omega^2-i\gamma\omega},
\label{nw}
\end{equation}
where $\omega_0$ is the resonance frequency, $f$ is the oscillator strength and $\gamma$ is the damping rate, e.g., of a dye molecule. $\epsilon_{c}$ is an arbitrary dielectric constant for the background polarization. 
Here, also throughout the text, we choose the positive imaginary part root for the refractive index in $n^2(\omega)=\epsilon(\omega)$, i.e., $n_I(\omega)>0$ with $n(\omega)=n_R(\omega)+i n_I(\omega)$. This choice constraints an absorbing (passive) medium. Actually, for a Lorentzian medium the two choices $n_I(\omega)>0$ and $n_R(\omega)>0$ are equivalent. In an electrically induced transparency~(EIT)-like medium~\cite{ScullyZubairyBook}, e.g., for the index enhancement phenomenon $n_R(\omega)>0$ is constrained~\cite{fleischhauer1992resonantly,fleischhauer2005electromagnetically,ScullyZubairyBook,panahpour2019refraction,gunay2020continuously,yuce2020ultra}. Because an auxiliary pulse is used in such systems which makes the system an active medium, thus, allowing $n_I(\omega)<0$. The choice $n_R(\omega)>0$, in Refs.~\cite{fleischhauer1992resonantly,fleischhauer2005electromagnetically,ScullyZubairyBook,panahpour2019refraction,gunay2020continuously,yuce2020ultra}, is made to avoid a negative-index medium. In all of the systems we study here, including the form of the hamiltonian in optomechanics, there is no gain. So we set the constraint $n_I(\omega)>0$. 

\begin{figure}
\begin{center}
\includegraphics[width=0.5\textwidth]{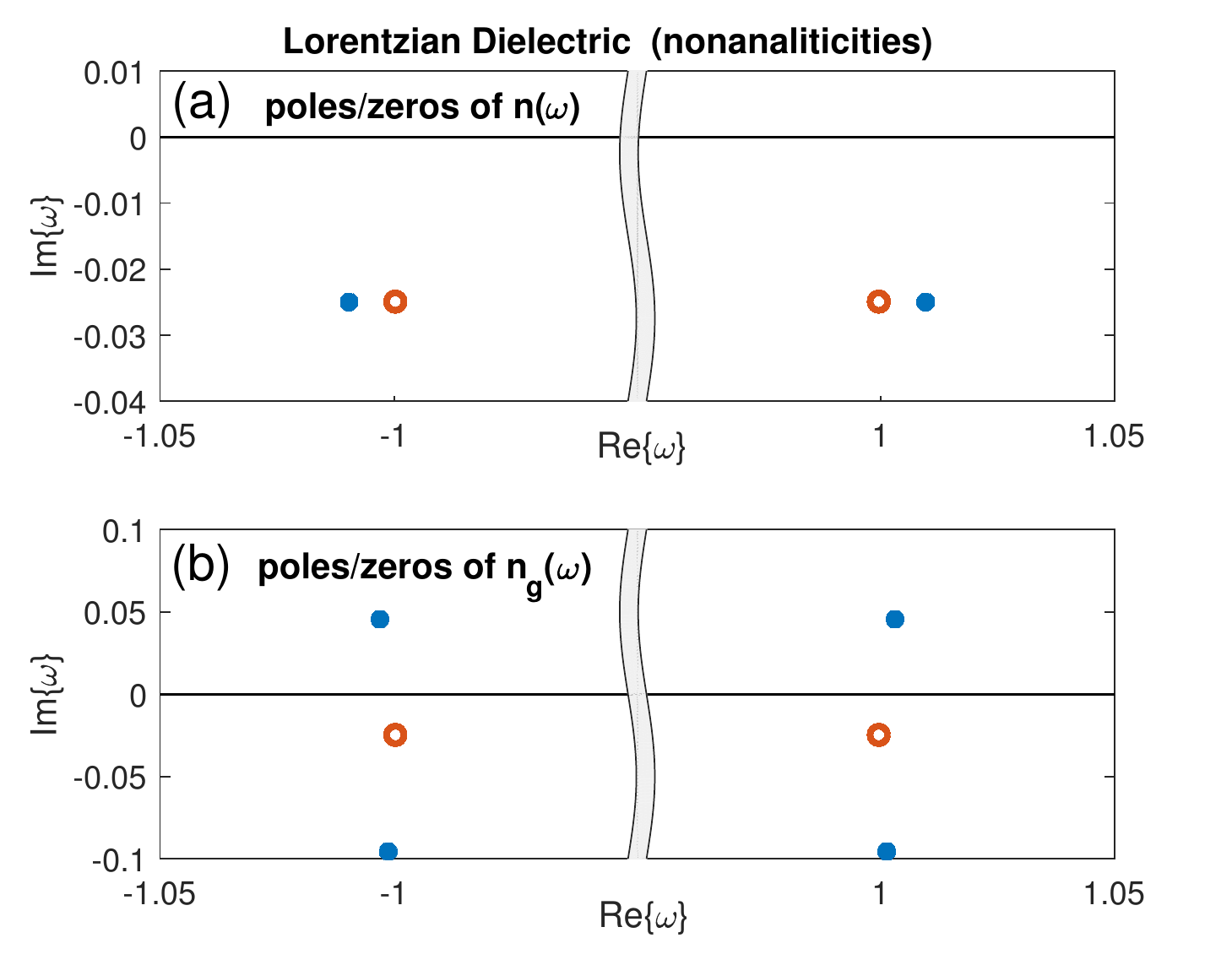}
\caption{(a) Nonalayticities of a Lorentzian refractive index $n(\omega)$, Eq.~(\ref{nw}), in the complex frequency plane. Circles (stars) are poles (zeros) of the refractive index $n(\omega)$. All poles and zeros rely in the lower-part of the complex frequency plane implying that $n(\omega)$ satisfies the Kramers-Kronig relations~(KKRs). (b) Nonalayticities of the group index $n_g(\omega)=n(\omega)+\omega dn(\omega)/d\omega$ for a Lorentzian dielectric function. The two zeros rely on the upper half of the CFP, implying a possible violation of the KKRs. Propagation velocity of a wavepacket, which is observed to be superluminal, is governed by the group index.}
\label{fig1}
\end{center}
\end{figure}

As it is well-known from electromagnetism text books~\cite{Jackson_book,griffiths_book} all the nonanalyticities of a Lorentzian index (more generally response function) are in the lower half of the complex frequency plane~(CFP), see Fig.~\ref{fig1}a. This is constrained by the principle that only the events from the past can affect the present fields/polarization $D(t)=\int_{-\infty}^t G(t-t') \: E(t')$. Thus $G(t-t')=0$ for $t'>t$. This implies a vanishing contour-integral in the upper half~(UH) of the CFP~(UH-CFP), thus, absence of nonanalyticities in the UH-CFP.

Wave packet propagation, e.g., movement of the pulse-center of energy/Poynting~\cite{peatross2000average,nanda2009superluminal,talukder2014direct,kohmoto2005nonadherence}, however, is governed by the group response (group velocity). That is, pulse-center movement~\cite{ChuPRL1982SL,WangNature200SL,chiao1999tunneling,talukder2005measurement} is shown to follow the behavior of the group velocity~\cite{peatross2000average,nanda2009superluminal,talukder2014direct,kohmoto2005nonadherence,tasginPRA2012testing,tasginBAJECE2013} which can be deduced from the envelop propagation in Eq.~(\ref{vg_envelop}).  Both pulse-center movement and group velocity demonstrate superluminal propagation in the experimental and theoretical studies~\cite{ChuPRL1982SL,WangNature200SL,chiao1999tunneling,talukder2005measurement,peatross2000average,nanda2009superluminal,talukder2014direct,kohmoto2005nonadherence,tasginPRA2012testing,tasginBAJECE2013}
.

Fig.~\ref{fig1}b shows that the group velocity $v_g=d\omega/dk$, or the group index $n_g(\omega)=n(\omega)+\omega \: dn/d\omega$, can possess nonanalyticities in the UH-CFP. Thus, this creates a possibility for the superluminal propagation in the second-order response, i.e., group velocity or pulse-center velocity. Because presence of an actual superluminal propagation necessitates the violation of KKRs via allowing the communication of causally-not-connected regions in classical electromagnetism. This result, i.e., presence of nonclassicalities of $n_g(\omega)$ in the UH-CFP, actually, is not so surprising and it is not a strike against the causal structure of the electromagnetism. Because the physical meaning of the pulse-center propagation is already under debate and actual signal can be demonstrated as the movement of an infinitesimally sudden disturbance~\cite{mandelstam1971lectures,stenner2003speed,shore2007superluminality,zhu2003propagation}. 

Therefore, the result is to be interpreted as follows. If $v_g(\omega)$, or pulse-center velocity, would correspond to an actual flow ---see Refs.~\cite{tasginPRA2012testing,talukder2014direct} for a counter demonstration--- it would allow communications between not-connected electromagnetic regions. 

\begin{figure}%[H]
\begin{center}
\includegraphics[width=0.5\textwidth]{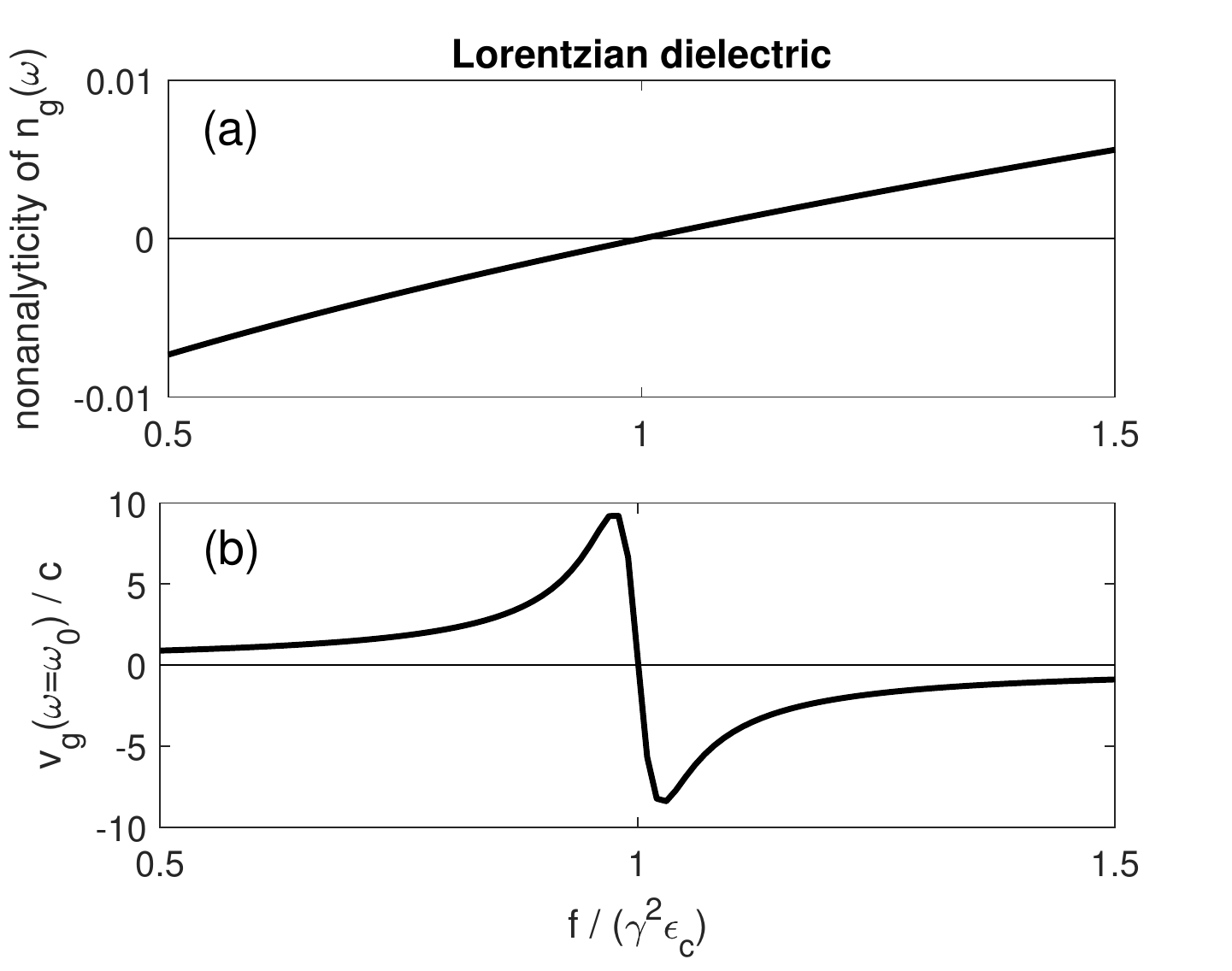}
\caption{ (a) The location of the nonanalyticities of the group index $n_g(\omega)$ for a Lorentzian dielectric, Eq.~(\ref{nw}). Nonanalyticities move into the upper half~(UH) of the complex frequency plane~(CFP), UH-CFP, for the oscillator strength $f>\gamma^2\epsilon$. $\gamma$ is scaled with the resonance frequency $\omega_0$. (b) The group velocity becomes negative after the same strength,  $f>\gamma^2\epsilon$, where the nonanalyticities in the second-order response $n_g(\omega)$ move into the UH-CFP.
  }
\label{fig2}
\end{center}
\end{figure}

The first time we obtained the result of Fig.~\ref{fig1}b, we instantly thought that we became able to explain the appearance of the superluminal group velocity $v_g>c$ in theoretical~\cite{peatross2000average,nanda2009superluminal,talukder2014direct,kohmoto2005nonadherence,tasginPRA2012testing,tasginBAJECE2013}  and experimental~\cite{ChuPRL1982SL,WangNature200SL,chiao1999tunneling,talukder2005measurement} studies. We, however, failed in that. Fig.~\ref{fig2} demonstrates that nonanalyticities of $n_g(\omega)$ move to the UH-CFP when the group velocity becomes negative $v_g<0$. For the parameters where $v_g>c$ takes place, the nonanalyticities of $n_g(\omega)$ remain in the lower half of the CFP. In other words, when $v_g>c$, the second-order response function $n_g(\omega)$ does not violate the KKRs. 

This result makes us consider the superluminal group behavior $v_g>c$ as being similar (or analogous) to the behavior of the phase velocity $v=c/n(\omega)>c$ for $n(\omega)<1$. 

More importantly, the observation ``$v_g<0$ when nonanalyticities move into the UH-CFP" intrigued us for further research. The phenomenon depicted in Fig.~\ref{fig2} appears in the second-order (group) response. The new question became: whether a similar situation appears also in the first-order response? In other words, does a negative phase velocity accompanies the presence of nonanalyticities of $R(\omega)$ in the UH-CFP, calculated in Refs.~\cite{wang2002causal,beck1991group,stern2012transmission,wangPRA2007theoretical,wang2016counterintuitive} ? In Sec.~\ref{sec:first-order}, we see that this indeed is the case. Both ({\sf 1}) interference effects, e.g., a jump-like behavior in a zero-amplitude finite spatial region~\footnote{\label{fn:jump} More explicitly, we imply the following. Let us consider an optical element divided into three regions of different refractive indices in space, e.g., like the one in Figs.~\ref{fig3} or \ref{fig8}. At specific wavelengths the reflected wave from the second interface can cancel the transmitted wave from the first interface perfectly. This may introduce a finite-size spatial region in which the wave at that specific frequency cannot enter for a specific wavelength. So, such a wave needs to \textit{jump}, e.g., from $x=0$ to $x=L$. This instance is possible to be responsible for the nonanalytical behavior we observe in the UH-CFP.   }, and ({\sf 2}) mirror-cavity coupling introduce nonanalyticities in the UH-CFP below/above critical vales of the system parameters. The transition into the UH-CFP is accompanied by a sudden change in the sign of the phase velocity~${}^{\ref{fn:causal}}$. We demonstrate the accompaniment also using the effective index method.

One critical conclusion we deduce from this subsection is: the different natures of two superluminal behaviors, $v_g>c$ and $v_g<0$, both referred as superluminal in the literature~\cite{ChuPRL1982SL,WangNature200SL,wang2006superluminal,tarhan2013superluminal}. The pulse-center movement for $v_g<0$~\cite{wang2006superluminal,tarhan2013superluminal}, not a physical propagation~\cite{mandelstam1971lectures,stenner2003speed,shore2007superluminality,zhu2003propagation}, violates the KKRs while $v_g>c$ does not.

%%%%%%%%%%%%%%%%%%%%%%%%%%%%%%%%%%%%%%%%%%%%%%%%%%%%%%%%%%%%%%%%%%%%%%%%%%%%%%%%%%%%%%%%%%%%%%%%%%%%%%%%%%%%%%%%%%%%%%%%%%%%%%%%%%%%%%%%%%%%%%%%%%%%
\subsection{Absorbing slab}  \label{sec:absorbing-slab}

An absorbing slab is also known for exhibiting negative group delays $\tau_R$ (second-order response) in the reflected wave~\cite{wang2006superluminal}. Ref.~\cite{wang2006superluminal} demonstrates that such a setup exhibits negative group delays $\tau_R$ at some specific wavelengths. Recent studies~\cite{Zubairy2014counterintuitive,wang2016counterintuitive}, however, find that the response functions $R(\omega)$ and $T(\omega)$ do not have any nonanalyticities in the UH-CFP, despite the observed negative group delays.

\begin{figure}%[H]
\begin{center}
\includegraphics[width=0.45\textwidth]{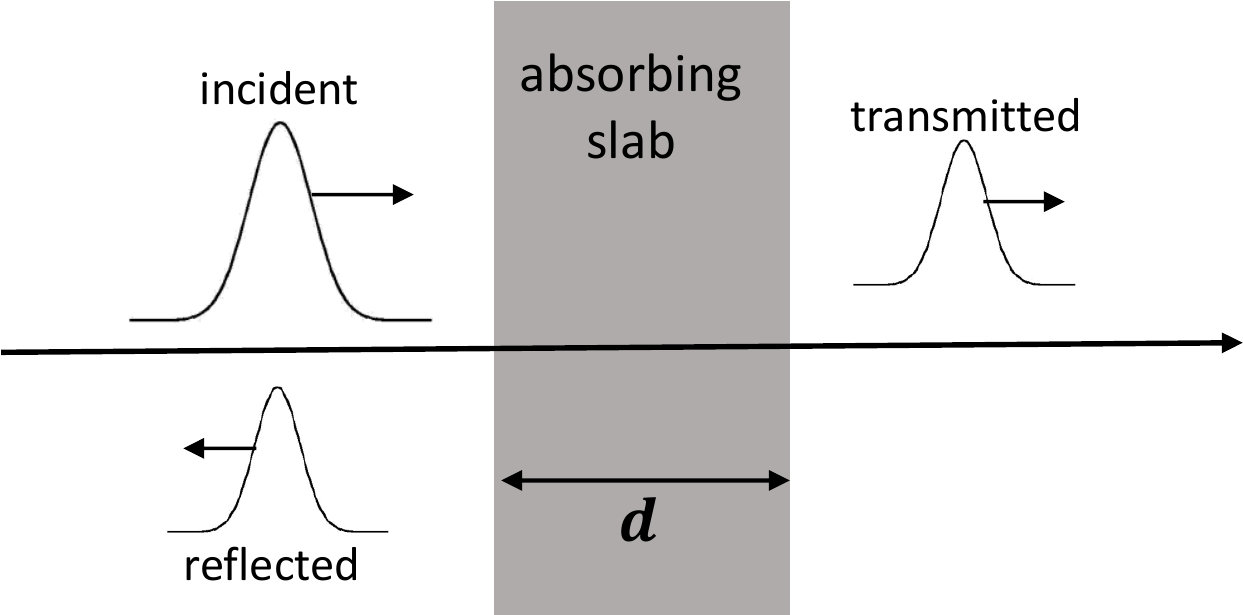}
\caption{Reflection and transmission through an absorbing slab. Group delay $\tau_R(\omega)$ of the reflected wave displays negative group velocities~\cite{wang2006superluminal}.
}
\label{fig3}
\end{center}
\end{figure}

\begin{figure}%[H]
\begin{center}
\includegraphics[width=0.5\textwidth, trim=0 0.7cm 0 0, clip]{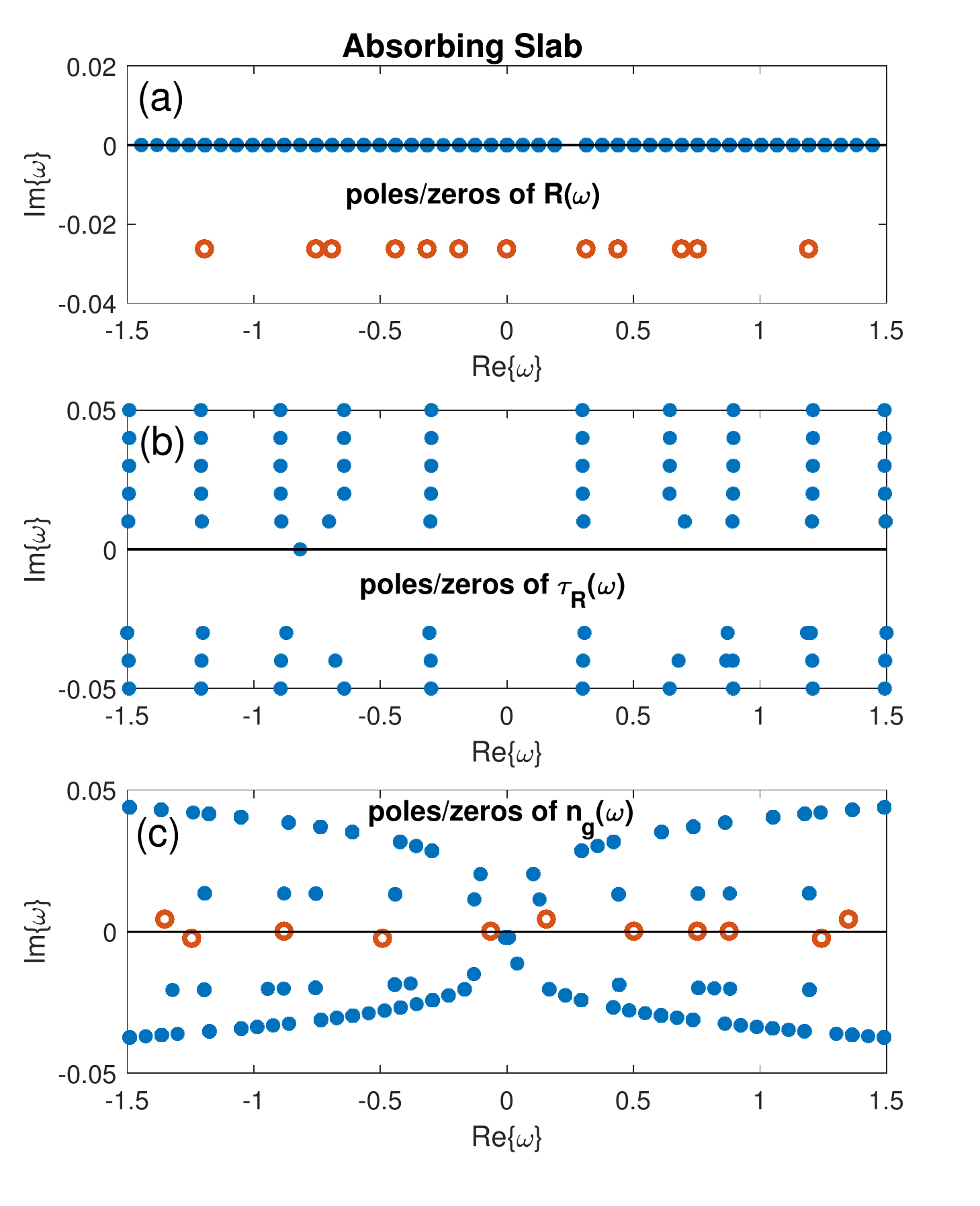}
%[trim=left bottom right top, clip]{
\caption{Absorbing slab. (a) Nonanalyticities of the first-order response function $R(\omega)$  do not rely in the upper-half~(UH) of the complex frequency plane~(CFP), UH-CFP. Neither a negative phase-velocity appears in the first-order response. However, nonanalyticities of the second-order (group) response (b) $\tau_R(\omega)$ and (c) $n_g^{\rm (eff)}(\omega)$ rely in the UH-CFP where negative group delays (velocities) are observed in Ref.~\cite{wang2006superluminal}.
}
\label{fig4}
\end{center}
\end{figure}

Noting that negative superluminal delay is calculated for the ``group" behavior, we also check the nonanalyticities of $n_g(\omega)$, or $\tau_R(\omega)$ equivalently. Since the optical setup in Fig.~\ref{fig3} is not a uniform medium, unlike in Sec.~\ref{sec:Lorentzian}, we calculate the effective group index $n_g^{(\rm eff)}(\omega)=n_{\rm eff}+\omega\: dn_{\rm eff}/d\omega$ from the effective index $n_{\rm eff}(\omega)$ as described in Sec.~\ref{sec:effective_index}. In Fig.~\ref{fig4}, we observe that: while the first-order [$R(\omega)$,$T(\omega)$] response does not exhibit any nonanalyticity in the UH-CFP, see Fig.~\ref{fig4}a, the second-order (group)  response displays nonanalyticities in the UH-CFP, see Fig.~\ref{fig4}b for $\tau_R(\omega)$ and Fig.~\ref{fig4}c for $n_g^{\rm (eff)}(\omega)$. In Figs.~\ref{fig4}b and \ref{fig4}c we double-check the presence of the nonanalyticities in the UH-CFP by evaluating $\tau_R(\omega)=0$, $1/\tau_R(\omega)=0$ and  $n_g^{\rm (eff)}(\omega)=0$, $1/n_g^{\rm (eff)}(\omega)=0$, respectively. We note that some of the nonanalyticities are missing in Fig.~\ref{fig4} due to our finite scanning interval for the solutions.

Therefore, one more time, we come to show the following phenomenon. A negative superluminal ``group" delay, governing a wave packet propagation, accompanies the presence of nonanalyticities in the UH-CFP for $\tau_R(\omega)=d\phi_R/d\omega$ and $n_g^{\rm (eff)}(\omega)$. $\phi_R(\omega)$ is the phase of the reflected wave $R(\omega)=|R(\omega)| e^{i\phi_R(\omega)}$. We still remark that group propagation is demonstrated not to correspond to a superluminal flow of the information~\cite{mandelstam1971lectures,stenner2003speed,shore2007superluminality,zhu2003propagation}, but it stands for the pulse-center propagation~\cite{peatross2000average,nanda2009superluminal,talukder2014direct,boyd2009controlling}. Nevertheless, if it has been corresponding to an information flow, presence of the nonanalyticities in the UH-CFP could have made the setup a ``possible" candidate for achieving superluminal response.

%%%%%%%%%%%%%%%%%%%%%%%%%%%%%%%%%%%%%%%%%%%%%%%%%%%%%%%%%%%%%%%%%%%%%%%%%%%%%%%%%%%%%%%%%%%%%%%%%%%%%%%%%%%%%%%%%%%%%%%%%%%%%%%%%%%%%%%%%%%%%%%%%%%%
\subsection{Reflection from a periodic structure} \label{sec:periodic}

\begin{figure}
\begin{center}
\includegraphics[width=0.45\textwidth]{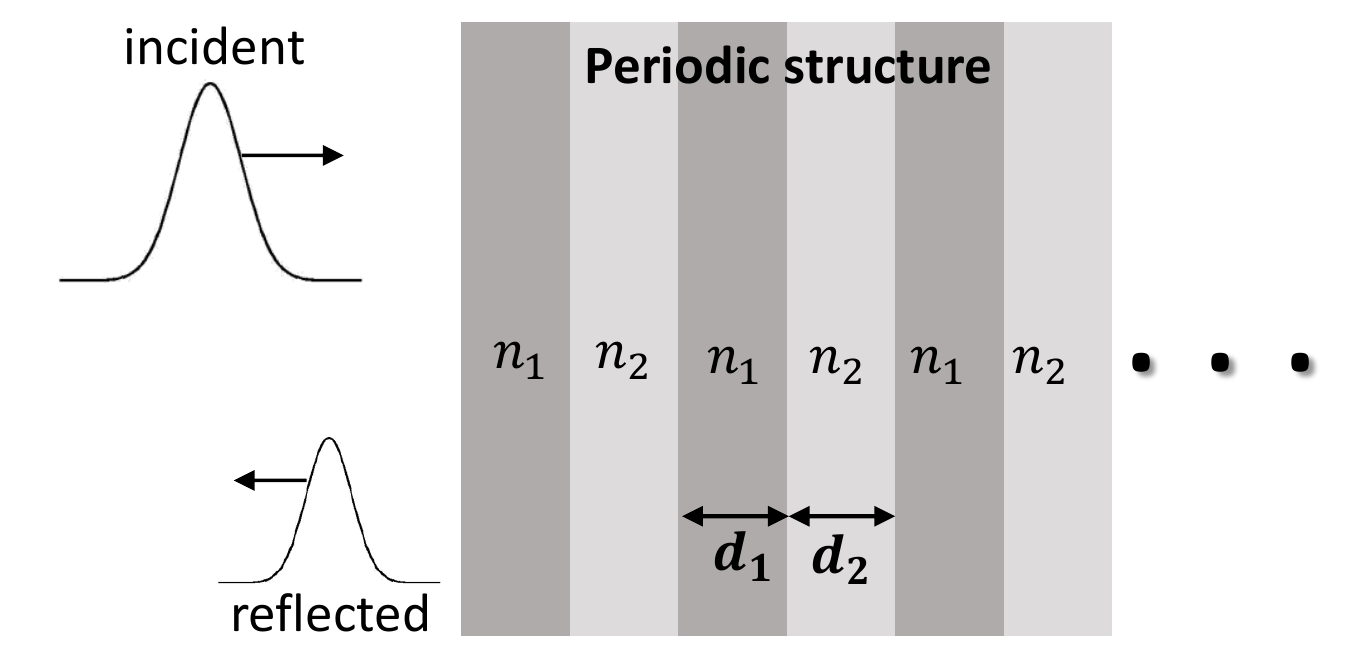}
\caption{A pulse reflected from a semi-infinite periodic structure exhibits negative group delays~\cite{levine1966reflection,lytvynenko2009wave,hwang2006correlation}, i.e., $\tau_R<0$. }
\label{fig5}
\end{center}
\end{figure}

\begin{figure}
\begin{center}
\includegraphics[width=0.5\textwidth]{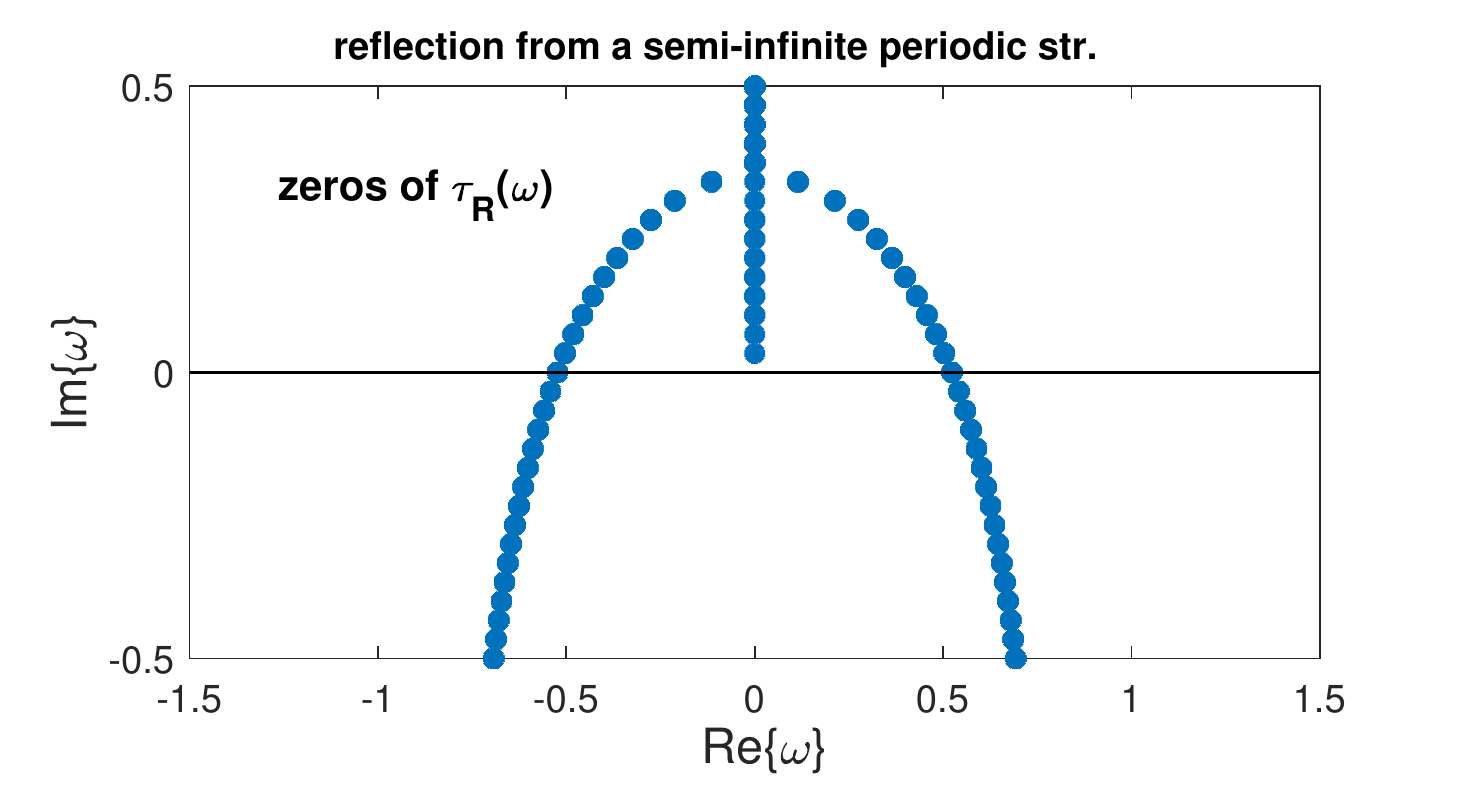}
\caption{Reflection from a periodic structure. Nonanalyticities of the second-order (group) response, i.e., $\tau_R(\omega)$, are located in the UH-CFP. Thus, negative group delays~\cite{levine1966reflection,lytvynenko2009wave,hwang2006correlation} accompany the violation of KKRs by the second-order response function $\tau_R(\omega)$. Zeros of the first-order response, not depicted, are on the real-$\omega$ axis. }
\label{fig6}
\end{center}
\end{figure}

One another optical setup known for displaying negative superluminal group delay is the reflection from a semi-infinite periodic structure~\cite{levine1966reflection,lytvynenko2009wave,hwang2006correlation} depicted in Fig.~\ref{fig5}. We show that, in this case too, the negative group delay~\cite{levine1966reflection,lytvynenko2009wave,hwang2006correlation} accompanies the presence of nonanalyticities in the UH-CFP belonging to the group response $\tau_R(\omega)$, see Fig.~\ref{fig6}. Such a behavior, again, appears in the second-order response (group behavior), i.e., $\tau_R(\omega)=d\phi_R / d\omega$. The nonanalyticities of the first-order [phase velocity or $R(\omega)$] response do not exhibit any nonanalyticity in the UH-CFP (not depicted).

We calculate an effective index $n_{\rm eff}(\omega)$ also for this setup as an alternative method for a double-check. However, we unable to make the subroutine solve the equations $n_g^{\rm (eff)}(\omega)=0$ or $1/n_g^{\rm (eff)}(\omega)=0$ for this setup.

It is worth further mentioning that: we also investigate the nonanalyticities of an infinite periodic structure~\cite{hwang2012periodic,morozov2011light,lalanne1996effective}. Effective index~\cite{hwang2012periodic} does not display any nonanalyticity in the UH-CFP.

%
%We consider a semi-infinite periodic structure and calculate the nonanalyticities of the reflection coefficient $R(\omega)$. We observe that all the zeros of $R(\omega)$ are on the real axis. We also calculate the nonanalyticities of the group delay $\tau_R(\omega)=\partial \phi_r/\partial \omega$ in Fig.~\ref{fig4} and obtain nnonanalyticies in the upper half of the CFP.

%%%%%%%%%%%%%%%%%%%%%%%%%%%%%%%%%%%%%%%%%%%%%%%%%%%%%%%%%%%%%%%%%%%%%%%%%%%%%%%%%%%%%%%%%%%%%%%%%%%%%%%%%%%%%%%%%%%%%%%%%%%%%%%%%%%%%%%%%%%%%%%%%%%%
%%%%%%%%%%%%%%%%%%%%%%%%%%%%%%%%%%%%%%%%%%%%%%%%%%%%%%%%%%%%%%%%%%%%%%%%%%%%%%%%%%%%%%%%%%%%%%%%%%%%%%%%%%%%%%%%%%%%%%%%%%%%%%%%%%%%%%%%%%%%%%%%%%%%
%%%%%%%%%%%%%%%%%%%%%%%%%%%%%%%%%%%%%%%%%%%%%%%%%%%%%%%%%%%%%%%%%%%%%%%%%%%%%%%%%%%%%%%%%%%%%%%%%%%%%%%%%%%%%%%%%%%%%%%%%%%%%%%%%%%%%%%%%%%%%%%%%%%%
\section{Nonanalyticities in the First-Order Response} \label{sec:first-order}

In the previous section we examined the accompaniment of a negative group velocity to the presence of group response nonanalyticities in the UH-CFP. We demonstrated this phenomenon in the second-order (group) response, i.e., where response function is associated with the derivative of the first-order response, $n_g(\omega)=n(\omega)+\omega\: dn/d\omega$ or $\tau_R(\omega)=d\phi_R/d\omega$.

In this section, we investigate if a similar accompaniment appears also in the first-order response, i.e., $n(\omega)$ or $R(\omega)$, $T(\omega)$. In difference to the examples studied above, in this section we present two optical setups in which locations of the nonanalyticities move from lower half of the CFP~(LH-CFP) to the upper half of the CFP~(UH-CFP) via change of parameters. In the first setup, Otto configuration~\cite{wang2016counterintuitive}, nonanalyticities appear in the UH-CFP due to interference effects~${}^{\ref{fn:jump}}$. In the second setup, an optomechanical system~\cite{agarwal2010electromagnetically,tarhan2013superluminal,genes2008robust,vitaliPRL2007optomechanical}, the mechanism responsible for the presence of nonanalyticities in the UH-CFP, however, has different origins. It is not the interference, but, achieving a critical cavity-mirror coupling strength onsetting correlations with a mechanical oscillator.

%%%%%%%%%%%%%%%%%%%%%%%%%%%%%%%%%%%%%%%%%%%%%%%%%%%%%%%%%%%%%%%%%%%%%%%%%%%%%%%%%%%%%%%%%%%%%%%%%%%%%%%%%%%%%%%%%%%%%%%%%%%%%%%%%%%%%%%%%%%%%%%%%%%%
\subsection{Otto configuration} \label{sec:Otto}

\subsubsection{Otto Configuration Setup}

\begin{figure}
\begin{center}
\includegraphics[width=0.5\textwidth]{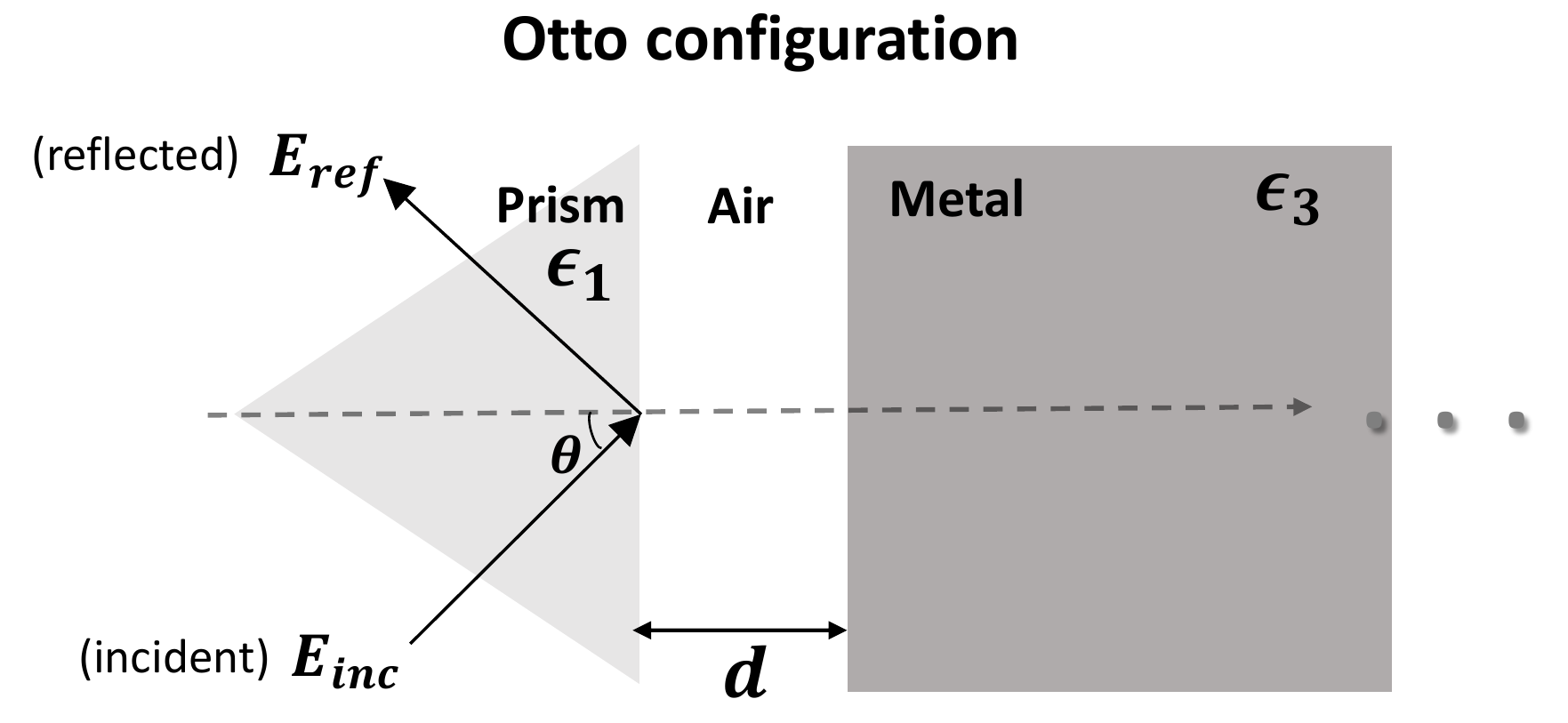}
\caption{Otto configuration; a 3 layer system. Optical light, incident to an air gap ($\epsilon_2=1$) from the a higher index prism ($\epsilon_1=9$), is reflected $E_{ref}$ and transmitted $E_{trans}$ into the air gap. The part transmitted into the air gap is reflected/transmitted again in a air/metal interface (not shown). Metal slab is very thick, thus, considered as semi-infinite in the calculations of Ref.~\cite{wang2016counterintuitive}. First-order response function $R(\omega)=E_{ref}/E_{inc}$ is investigated in the present work. $\theta=20^{o}$. 
}
\label{fig7}
\end{center}
\end{figure}

In Fig.~\ref{fig7}, we depict an Otto configuration~\cite{wang2016counterintuitive}. An optical frequency light $E_{inc}$ is incident to a prism/air gap interface at an angle $\theta=20^o$~\cite{wang2016counterintuitive}. Part of the wave is transmitted into the air gap of thickness $d$ and remaining part is reflected $E_{ref}$. A very thick metal (in general an absorbing medium) slab of dielectric function $\epsilon_3$ follows the air gap, where again a transmission/reflection takes place. Ref.~\cite{wang2016counterintuitive} investigates the locations of the nonanalyticities  of the transfer function (reflection) $R(\omega)=E_{ref}/E_{inc}$. The reflection coefficient are calculated~\cite{wang2016counterintuitive} as
\begin{equation}
\resizebox{.99\hsize}{!}{$
R(\omega)=\frac{ (p_1+p_2)(p_2-p_3)e^{-k\eta_a d} + (p_1-p_2)(p_2+p_3)e^{k\eta_a d}  }{ (p_1-p_2)(p_2-p_3)e^{-k\eta_a d} + (p_1+p_2)(p_2+p_3)e^{k\eta_a d} }$
}
\label{rw_Otto}
\end{equation}
for an infinitely thick metal slab, where $k$ is the wavenumber in vacuum, $p_1=(\epsilon_1-k_y^2/k^2)^{1/2}/\epsilon_1$, $p_2=i\eta_a/\epsilon_2$, $p_3=i\eta_b /\epsilon_3$, $\eta_a=(k_y^2/k^2-\epsilon_2)^{1/2}$, $\eta_b=(k_y^2/k^2-\epsilon_3)^{1/2}$, $k_y=k\epsilon_1^{1/2}\sin\theta$~\footnote{Ref.~\cite{wang2016counterintuitive} calculates also the nonanalyticities of the transmission into the $\epsilon_3$ medium and obtains the same conditions for the locations of the nonanalyticities. Here, we do not consider it since $\epsilon_3$ is a semi-infinite medium. }. Ref~\cite{wang2016counterintuitive} calculates a 2D diagram of Re$\{\epsilon_3\}$ and Im$\{\epsilon_3\}$ axes demonstrating in which regimes nonanalyticities of the $R(\omega)$ are  in the UH-CFP or LH-CFP. Nonanalyticities are shown to appear in the UH-CFP also for a Drude dielectric function (for $\epsilon_3$).

\subsubsection{Nonanalyticities}

\begin{figure}
\begin{center}
\includegraphics[width=0.5\textwidth]{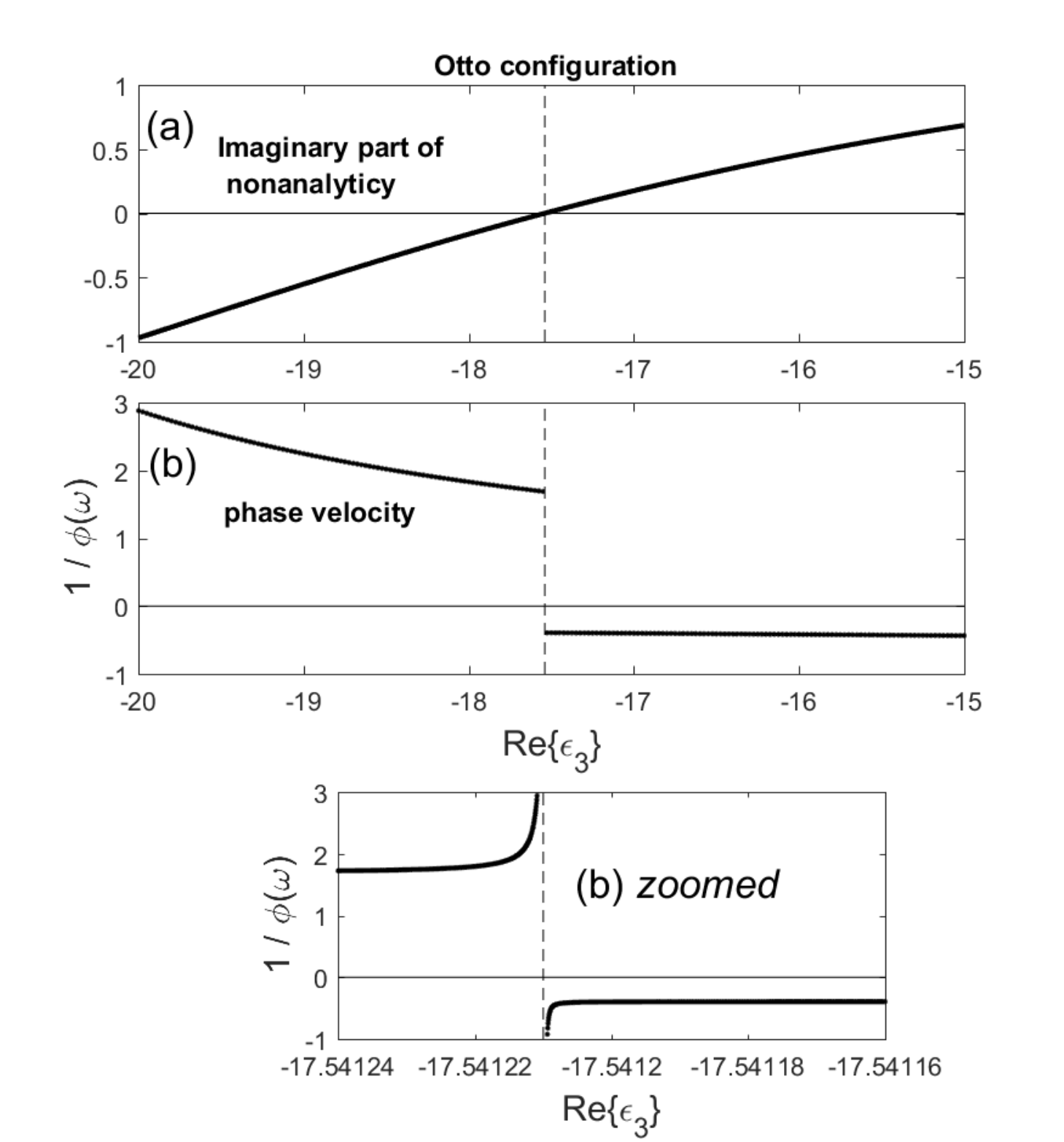}
\caption{A negative phase-velocity accompanies the movement of the nonanalyticities into the upper-half~(UH) of the complex frequency plane~(CFP), UH-CFP, also for the first-order response. (a) The location of the nonanalyticities of $R(\omega)$ in an Otto configuration (Fig.~\ref{fig7}). Nonanalyticities of the first-order response become located in the UH-CFP for Re$\{\epsilon_3\}>-$17.54. $\epsilon_3$ is the dielectric constant of the absorbing medium in Fig.~\ref{fig7} and we set Im$\{\epsilon_3\}=$5. (b) Exactly at the same place Re$\{\epsilon_3\}>-$17.54, the phase of the reflected wave, the sine term in Eq.~(\ref{tauR_envelop}), displays an abrupt change from positive to negative.
}
\label{fig8}
\end{center}
\end{figure}

In Fig.~\ref{fig8}a, we regenerate the results of Ref.~\cite{wang2016counterintuitive} for a fixed Im$\{\epsilon_3\}$=5 and when Re$\{\epsilon_3\}$ scans -20 and -15~\footnote{These two values correspond to points (b) and (c) in Ref.~\cite{wang2016counterintuitive} Fig.~2.}. Nonanalyticity of $R(\omega)$ move into the UH-CFP at about Re$\{\epsilon_3\}\simeq-$17.54 as demonstrated in Fig.~\ref{fig8}a and Ref.~\cite{wang2016counterintuitive}. In Fig.~\ref{fig8}b we demonstrate that the phase of the transfer function $R(\omega)=|R(\omega)| e^{i\phi_R}$ changes sign to a negative value at Re$\{\epsilon_3\}\simeq -$17.54 indicating a sign change in the phase velocity. This happens exactly at the same place where nonanalycities move to the UH-CFP in Fig.~\ref{fig8}a.

Therefore, we clearly observe that a negative phase velocity~\footnote{We kindly remark that $\phi_R$ in $R(\omega)=E_{ref}/E_{inc}=|R(\omega)|e^{i\phi_R}$ reports the change of the phase with respect to the incident wave.} accompanies the appearance of nonanalyticities in the UH-CFP \textit{also} in the ``first-order" response functions $R(\omega)$, $T(\omega)$. This is in analogy with the behavior of the second-order response functions studied in Sec.~\ref{sec:second-order}.

\subsubsection{Nonanalyticities via  effective index method}

 As an alternative method, for a double-check, we also investigate the behavior of the effective dielectric function $\epsilon_{\rm eff}(\omega)$~\cite{smith2002determination,yoo2019causal}. We check ({\sc 1}) if the nonanalyticities of $\epsilon_{\rm eff}(\omega)$ move into the UH-CFP at the same parameter with $R(\omega)$,i.e., as in Fig.~\ref{fig8}, and ({\sc 2}) if the effective index (velocity) display a sign change at the place Re$\{\epsilon_3\}\simeq -$17.54.

\begin{figure}
\begin{center}
\includegraphics[width=0.5\textwidth]{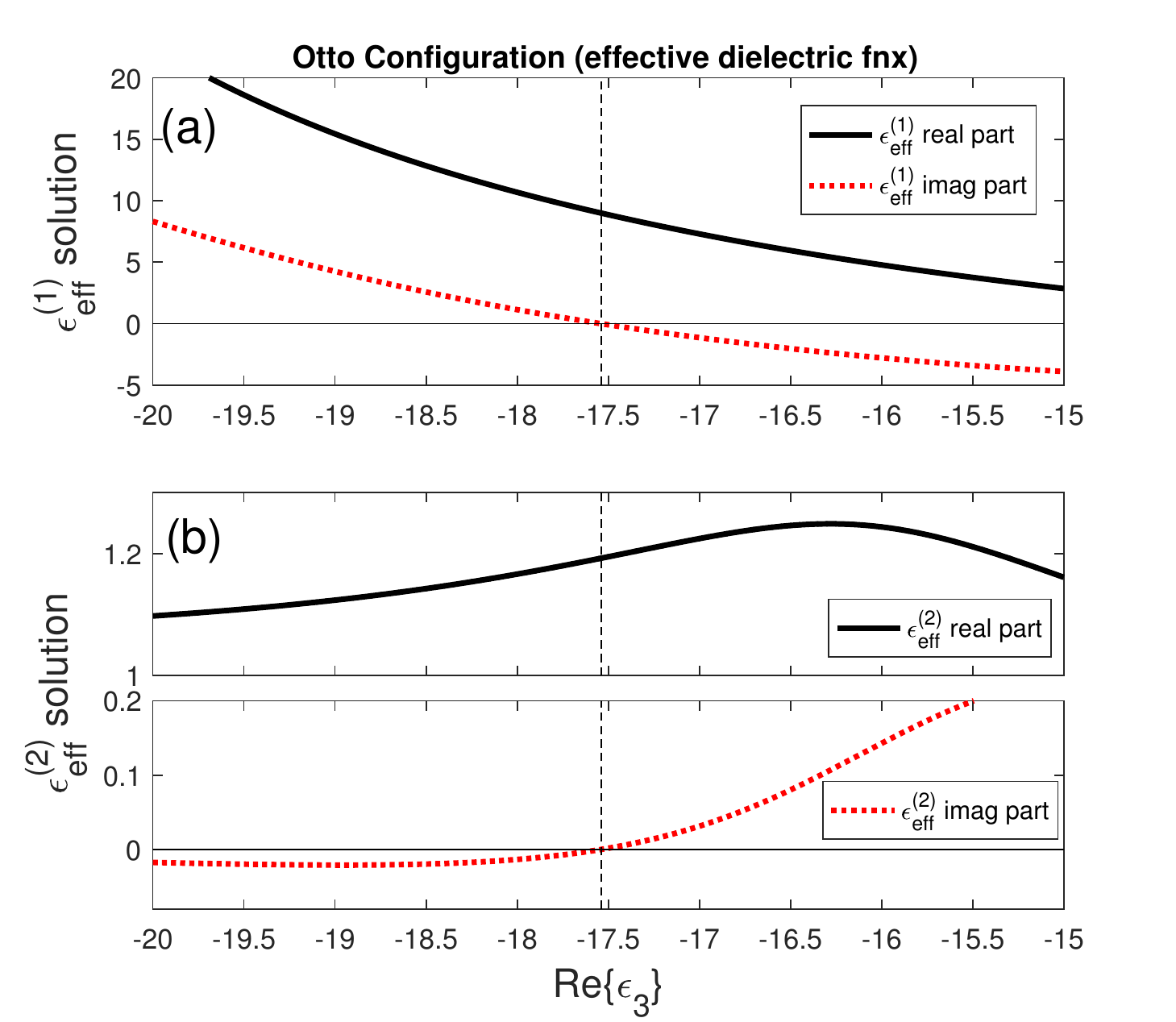}
\caption{The effective dielectric function $\epsilon_{\rm eff}(\omega)$ for the Otto configuration in Fig.~\ref{fig7}. The effective dielectric, i.e., obtained from Eq.~(\ref{rw_Rw}), has two solutions (a) $\epsilon_{\rm eff}^{(1)}$ and (b) $\epsilon_{\rm eff}^{(2)}$. Imaginary part of both solutions display a sign change at Re$\{\epsilon_3\}\simeq -$17.54 where ({\sf 1}) nonanalyticities move into the UH-CFP in Fig.~\ref{fig8}a and ({\sf 2}) phase-velocity jumps from a positive to a negative value in Fig.~\ref{fig8}b.
}
\label{fig9}
\end{center}
\end{figure}

Ref.~\cite{wang2016counterintuitive} calculates $R(\omega)$, given in Eq.~(\ref{rw_Otto}), for a semi-infinite metallic slab $\epsilon_3$ (we mention it as very thick)  ---i.e., not a finite region as in Fig.~\ref{fig13} or Fig.~\ref{fig3}. So, we calculate the effective dielectric function from the prism/air gap interface. Reflection of TM waves on an $n_1$/$n_2$ (or $n_2$/$n_{\rm eff}$) interface is given by~\cite{Jackson_book,griffiths_book}
\begin{equation}
r(\omega)= \frac{  \epsilon_{\rm eff}\cos\theta - n_1\sqrt{\epsilon_{\rm eff}-n_1^2\sin^2\theta}  }{  \epsilon_{\rm eff}\cos\theta + n_1\sqrt{\epsilon_{\rm eff}-n_1^2\sin^2\theta}  }  
\end{equation}
for a nonmagnetic medium. We determine the effective dielectric function $\epsilon_{\rm eff}=n_{\rm eff}^2$ by equating this reflection coefficient to the one calculated for the Otto configuration in Ref.~\cite{wang2016counterintuitive}, i.e., the one given in Eq.~(\ref{rw_Otto}),
\begin{eqnarray}
r(\omega)=R(\omega).
\label{rw_Rw}
\end{eqnarray}
Eq.~(\ref{rw_Rw}) has two solutions for $\epsilon_{\rm eff}(\omega)$, $\epsilon_{\rm eff}^{(1)}(\omega)$ and $\epsilon_{\rm eff}^{(2)}(\omega)$, as depicted in Fig.~\ref{fig9} for the same parameter set used in Fig.~\ref{fig8} and Ref.~\cite{wang2016counterintuitive}.

\begin{figure}
\begin{center}
\includegraphics[width=0.5\textwidth]{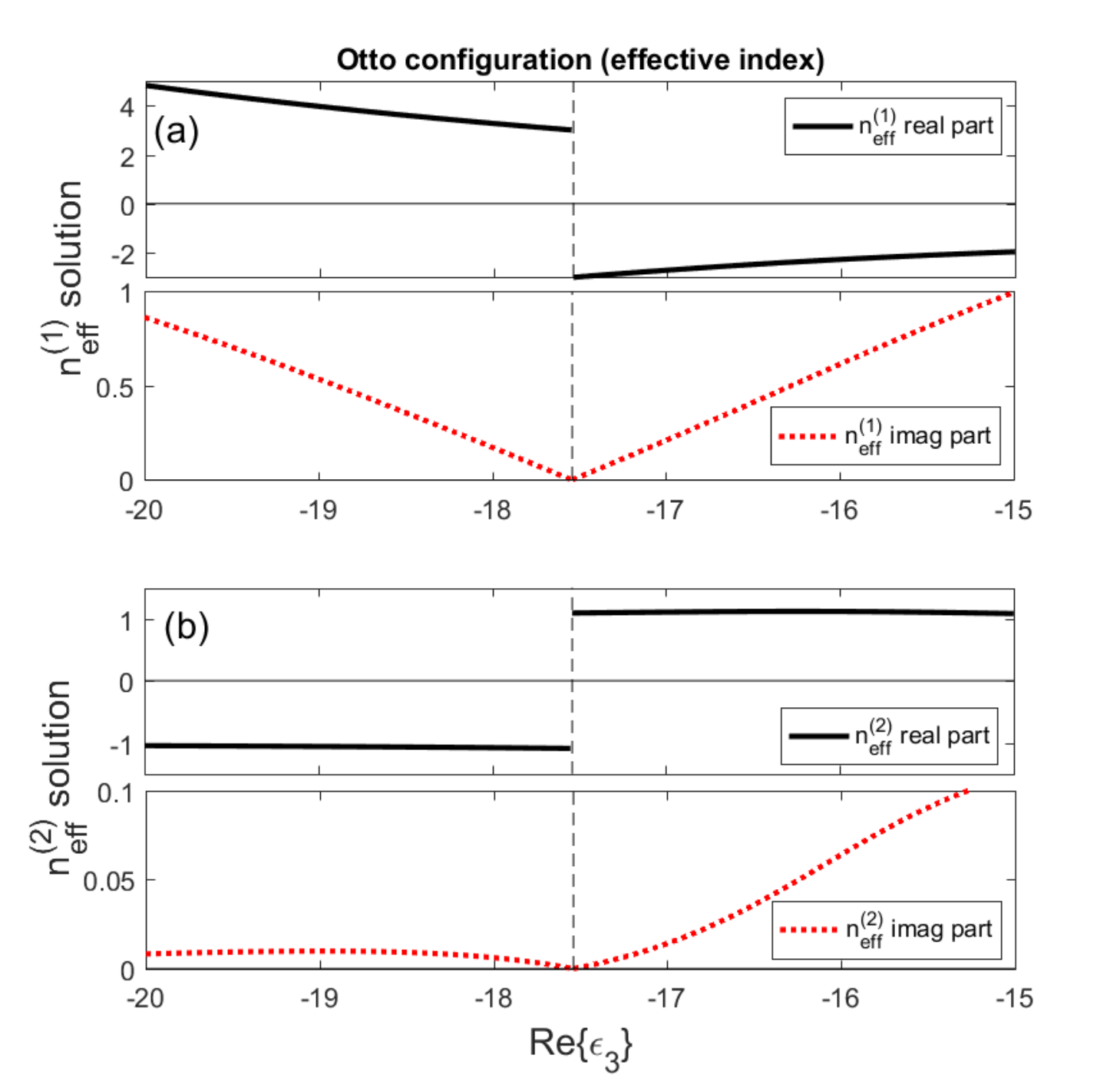}
\caption{ Effective indices $n_{\rm eff}^{(1,2)}$ corresponding to the effective dielectric solutions $\epsilon_{\rm eff}^{(1,2)}(\omega)$.  Im$\{n_{\rm eff}^{(1,2)}\}>0$ is constrained as the Otto configuration is a passive optics element. Both solutions $n_{\rm eff}^{(1,2)}$ display a sign change (i.e., real part) exactly at the same place Re$\{\epsilon_3\}\simeq -$17.54 where ({\sf 1}) nonanalyticities move into the UH-CFP in Fig.~\ref{fig8}a and ({\sf 2}) phase-velocity jumps from a positive to a negative value in Fig.~\ref{fig8}b. Sign change for both solutions obliges the presence of the negative index either for Re$\{\epsilon_3\} < -$17.54 or Re$\{\epsilon_3\} > -$17.54. The first solution can be decided easily by considering the sign change of the phase velocity in Fig.~\ref{fig8}b.
}
\label{fig10}
\end{center}
\end{figure}

We observe that imaginary part of both solutions $\epsilon_{\rm eff}^{(1,2)}(\omega)$ change sign exactly at  Re$\{\epsilon_3\}\simeq-$17.54 where nonclassicalities move into the UH-CFP in Fig.~\ref{fig8}a. In Fig.~\ref{fig10}, we also present the corresponding effective refractive indices $[n_{\rm eff}^{(1,2)}]^2=\epsilon_{\rm eff}^{(1,2)}$. In calculating $n_{\rm eff}^{(1,2)}$, we choose the Im$\{n_{\rm eff}^{(1,2)}\}>0$ roots. Because the system is impossible to display gain. Im$\{n_{\rm eff}^{(1,2)}\}>0$  as the setup contains only absorptive elements.

Fig.~\ref{fig10} demonstrates that $n_{\rm eff}^{(1)}(\omega)$ solution presents a sudden sign change, again exactly at Re$\{\epsilon_3\}\simeq -$17.54, where nonanalyticities move into the UH-CFP and transition to a negative phase-velocity~\footnote{We round up the critical value as Re$\{\epsilon_3\}\simeq>-$17.54. The two phenomena take place (accompany each other) at the same Re$\{\epsilon_3\}$ down to 6 digits which we do not present. } is observed in Fig.~\ref{fig8}. We indicate that both solutions $n_{\rm eff}^{(1,2)}(\omega)$ present a sudden sign change at Re$\{\epsilon_3\}\simeq -$17.54. Though their sign change is opposite, appearance of a sign change in both solutions obliges the presence of a negative effective index material in either sides, i.e., Re$\{\epsilon_3\}\simeq< -$17.54 or Re$\{\epsilon_3\}\simeq>-$17.54. Keeping in the view that phase velocity changes sign from positive to negative in Fig.~\ref{fig8}b, one can decide the choice of the $n_{\rm eff}^{(1)}(\omega)$ solution easily.

\begin{figure}
\begin{center}
\includegraphics[width=0.5\textwidth]{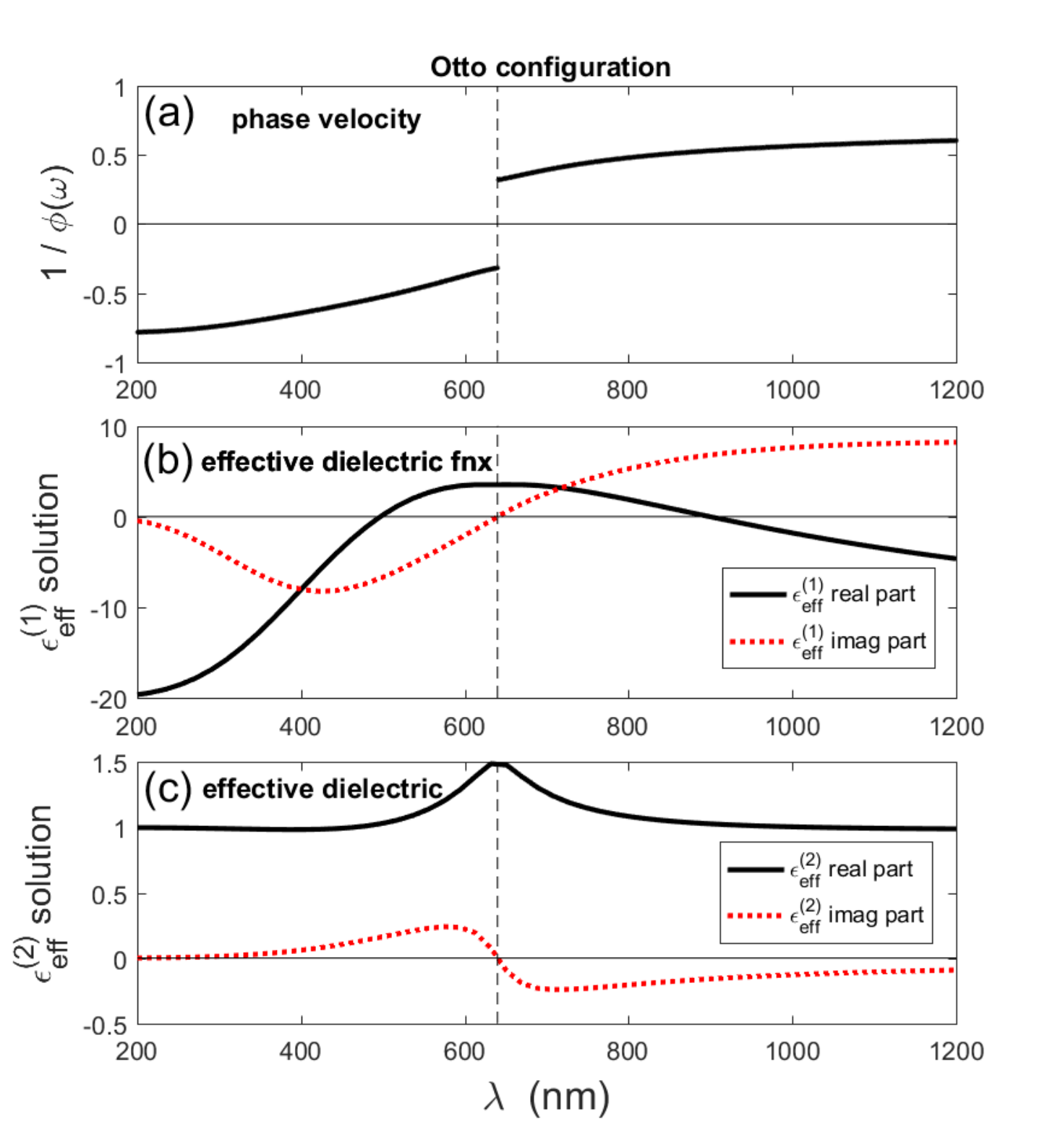}
\caption{Wavelength domain. Behaviors of the (a) phase-velocity and (b,c) effective dielectric functions $\epsilon_{\rm eff}^{(1,2)}(\omega)$ in the wavelength domain. All changes take place at the same wavelength $\lambda^*\simeq$ 640 nm.
 }
\label{fig11}
\end{center}
\end{figure}

\begin{figure}
\begin{center}
\includegraphics[width=0.5\textwidth]{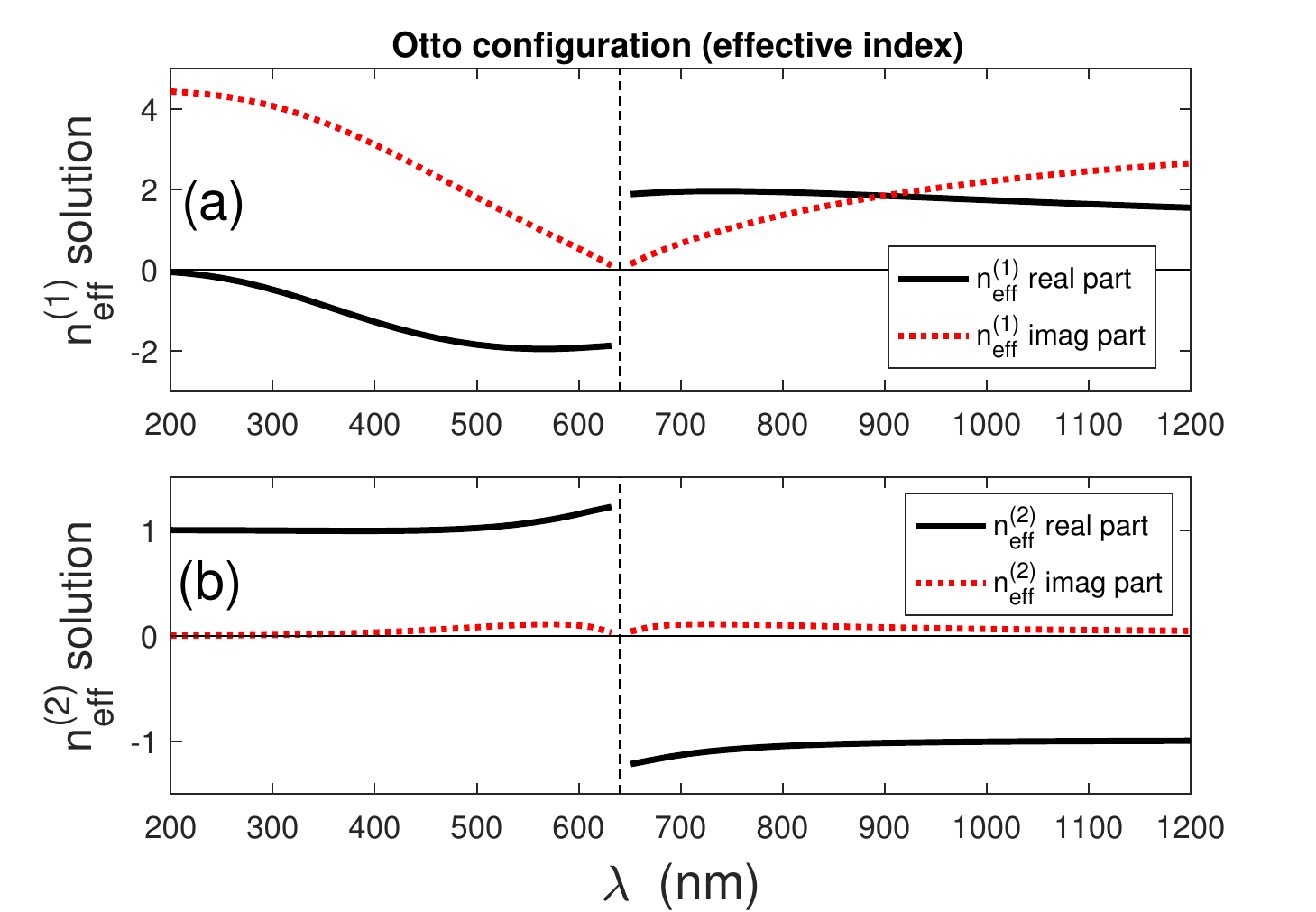}
\caption{Behaviors of effective indices $n_{\rm eff}^{(1,2)}(\omega)$ in the wavelength domain. (a) First solution $n_{\rm eff}^{(1)}(\omega)$ displays the same sign change behavior with the phase-velocity (Fig.~\ref{fig11}a). The sign change takes place exactly at the same wavelength $\lambda^*\simeq$640 nm. Appearance of sign changes on both solutions obliges the presence of a negative index. Examining the behavior of the phase-velocity in Fig.~\ref{fig11}a, one can easily decide the first solution $n_{\rm eff}^{(1)}(\omega)$.  
}
\label{fig12}
\end{center}
\end{figure}

In Figs.~\ref{fig11} and \ref{fig12}, we examine the phenomenon also in the wavelength domain. Fig.~\ref{fig11}a shows that phase velocity demonstrates negative values when $\lambda<\lambda_{\rm crt} \simeq $640 nm. At exactly the same wavelength, $\lambda_{\rm crt}$, effective index solutions $\epsilon_{\rm eff}^{(1,2)}(\omega)$ display a sign change in the imaginary part. In Fig.~\ref{fig12}a, we observe that the first solution for the effective index $n_{\rm eff}^{(1)}(\omega)$ displays the sign change observed in Fig.~\ref{fig11}a, i.e., for the phase of $R(\omega)$. This is consistent with Figs.~\ref{fig8}a,b and \ref{fig10}a. The second solution can be omitted via similar considerations mentioned in the previous paragraph.

{\it In summary}, in this section we show that a negative phase velocity accompanies the movement of the nonanalyticities to the UH-CFP in the first-order response, $R(\omega)$, of an Otto configuration. We demonstrate this phenomenon both ({\sc 1}) by exploring the phase of $R(\omega)$, i.e., $\phi_R(\omega)$, and ({\sc 2}) by exploring the sign of the refractive index we obtain via widely used effective index method~\cite{smith2002determination,yoo2019causal}. In both cases,  negative velocity or negative index onset exactly at the same critical parameter where nonanalyticities of $R(\omega)$ move into the UH-CFP.

%%%%%%%%%%%%%%%%%%%%%%%%%%%%%%%%%%%%%%%%%%%%%%%%%%%%%%%%%%%%%%%%%%%%%%%%%%%%%%%%%%%%%%%%%%%%%%%%%%%%%%%%%%%%%%%%%%%%%%%%%%%%%%%%%%%%%%%%%%%%%%%%%%%%
\subsection{Optomechanical System} \label{sec:optomechanics}

In this subsection, we investigate the nonanalyticities of an optomechanical system for the first time ---to our best knowledge. We determine the nonanalyticities of the first-order response functions $R(\omega)$ and $T(\omega)$. We demonstrate that nonanalyticities move into the UH-CFP above a critical cavity-mirror coupling $g>g_{\rm crt}$. This transition is shown to be accompanied by a sign change in the reflected phase, see Fig.~\ref{fig14}, similar to our findings in the Otto configuration~(Sec.~\ref{sec:Otto}). As a double-check, we show that the phase of the effective index also turns its sign from positive to negative exactly at the same critical coupling $g>g_{\rm crt}$, see Fig.~\ref{fig15}. We also show that: while an optomechanical system is a gain medium, the effective hamiltonian (responsible for the movement of the nonanalyticities to the UH-CFP) has a form which does not contain gain.

Presenting the crucial results of the subsection in advance, we now move to describing how we obtain them.

\subsubsection{Optomechanical system}

\begin{figure}
\begin{center}
\includegraphics[width=0.5\textwidth]{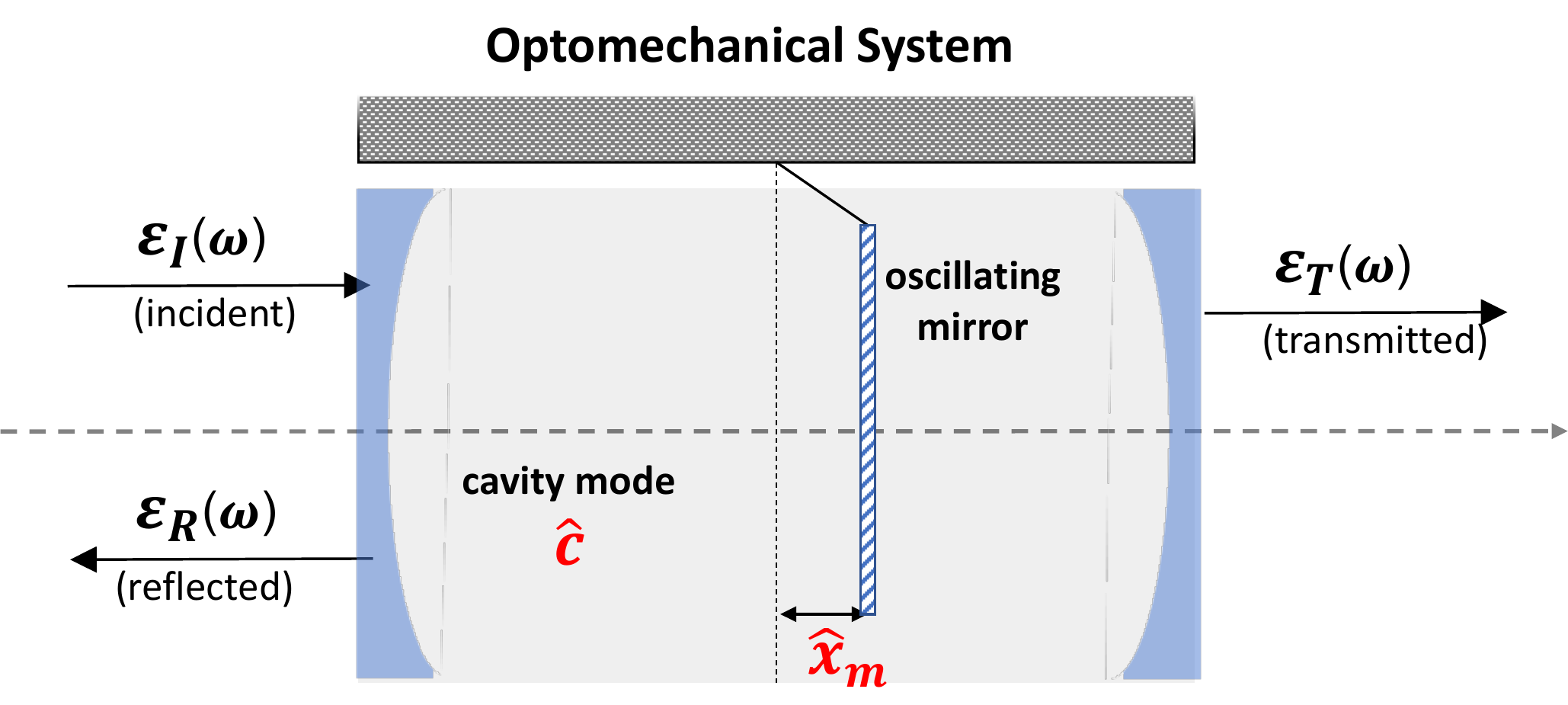}
\caption{ An optomechanical system. The cavity mode $\hat{c}$ is pumped by a strong coupler laser of frequency $\omega_{\rm L}$. The oscillating mirror, placed inside the cavity, interacts with cavity field via a radiation-pressure like coupling. The coupler laser is employed to turn up the effective cavity-mirror coupling.  The cavity-mirror coupling  introduces entanglement~\cite{genes2008robust,vitaliPRL2007optomechanical}, single-mode nonclassicality~(e.g., quadrature squeezing)~\cite{tasgin2019entanglement} and nonanalyticities [of $R(\omega)=\varepsilon_R/\varepsilon_I$] in the UH-CFP as we show in this study.
}
\label{fig13}
\end{center}
\end{figure}

{ An optomechancial system}~\cite{tarhan2013superluminal,agarwal2010electromagnetically,genes2008robust,vitaliPRL2007optomechanical}, depicted in Fig.~\ref{fig13}, consists of an optical cavity of resonance $\omega_c$, cavity mode operator $\hat{c}$, and an oscillating mirror of resonance $\omega_m\sim10^6$ Hz of mode operator $\hat{a}_m$. The cavity mode $\hat{c}$ interacts with the oscillating mirror via a radiation pressure type coupling $\hat{\cal H}_{\rm int}=\hbar g_0 \hat{c}^\dagger\hat{c}\hat{x}_m$ where $\hat{x}_m=(\hat{a}_m^\dagger+\hat{a}_m)/\sqrt{2}$ is the displacement of the mechanical oscillator from its equilibrium. A strong (coupler) laser of frequency $\omega_{\rm L}$ pumps the cavity in order to increase the effective coupling between the mirror and the cavity (so, it is called coupler laser). The full hamiltonian of the system, in the rotating frame with the laser frequency is given by~\cite{genes2008robust,vitaliPRL2007optomechanical}
\begin{equation}
\hat{\cal H}=\hbar \Delta_c \hat{c}^\dagger \hat{c} + \hbar \omega_m \hat{a}_m^\dagger \hat{a}_m + \hbar g_0 \hat{c}^\dagger \hat{c} \hat{a}_m + i\hbar \varepsilon_{\rm L} (\hat{c}^\dagger - \hat{c}), \label{hamiltonian_full}
\end{equation}
where $\Delta_c=\omega_c-\omega_{\rm L}$ and $g_0$ is the ``bare" cavity-mirror coupling strength. The last term governs the interaction of the cavity mode with the 
coupler laser of amplitude proportional to $\varepsilon_{\rm L}$.

Quantum optics features of such a system have been studied both in the second-quantized picture, investigating the cavity-mirror entanglement~\cite{genes2008robust,vitaliPRL2007optomechanical} and cavity mode nonclassicality~\cite{tasgin2019entanglement}, and in the first-quantized picture demonstrating the EIT-like behavior~\cite{tarhan2013superluminal,agarwal2010electromagnetically}. We initially keep things second-quantized in order to demonstrate the form of the effective hamiltonian Refs.~\cite{tarhan2013superluminal,agarwal2010electromagnetically,genes2008robust,vitaliPRL2007optomechanical} work around. Then, we work with an infinitesimally weak probe pulse to obtain the linear response~\cite{tarhan2013superluminal,agarwal2010electromagnetically} of the system.

We note that, we only re-present the derivations of Refs.~\cite{tarhan2013superluminal,agarwal2010electromagnetically,genes2008robust,vitaliPRL2007optomechanical}. That is, we do not present any new result except the locations for the first-order and second-order response functions of an optomechanical system.

Refs.~\cite{genes2008robust,vitaliPRL2007optomechanical} linearize the $\hat{\cal H}_{\rm int}=\hbar g_0 \hat{c}^\dagger\hat{c}\hat{x}_m$ about the steady-state value of the optomechanical system, i.e., $\hat{c}=\alpha_c + \delta\hat{c}$ and $\hat{a}_m=\alpha_m + \delta\hat{a}_m$. $\delta\hat{c}$ and $\delta\hat{a}_m$ are the quantum noise operators which by themselves determine the entanglement and nonclassicality features~\cite{simon1994quantum}. They represent merely the noise and have zero expectations, i.e., $\langle\delta\hat{c}\rangle=0$ and $\langle \delta\hat{a}_m\rangle=0$. The steady-state amplitudes are obtained from the steady-state of equations
\begin{eqnarray}
&&\dot{x}_m= \omega_m p_m, \label{langevin_clas_xm}
\\
&&\dot{p}_m=-\gamma_m p_m - \omega_m x_m + g_0 |\alpha_c|^2, \label{langevin_clas_pm}
\\
&&\dot{\alpha}_c= -(\kappa+i\Delta_c)\alpha_c +i g_0 x_m \alpha_c + \varepsilon_{\rm L}, \label{langevin_clas_alphac}
\end{eqnarray}
using the Heisenberg equations of motion, e.g., $i\hbar \dot{\hat{c}}=[\hat{c},\hat{\cal H}]$, by replacing operators by c-numbers~\cite{agarwal2010electromagnetically}, e.g., $\hat{c}\to\alpha_c$, and including the damping of the cavity $\gamma_c$ and the mirror $\gamma_m$ to the reservoirs. For double and single sided cavities, $\kappa=2\gamma_c$ and $\kappa=\gamma_c$, respectively. The steady-state values are obtained from Eqs.~(\ref{langevin_clas_xm})-(\ref{langevin_clas_alphac}) by setting $\dot{x}_m=\dot{p}_m=\dot{\alpha}_c=0$. They come out as $\bar{p}_m=0$, $\bar{x}_m=g_0/\omega_m|\alpha_c|^2$ and $\bar{\alpha}_c=\varepsilon_{\rm L}/[\kappa+i(\Delta_c-g_0 \bar{x}_m)]$~\cite{genes2008robust,vitaliPRL2007optomechanical}.

So, the effective interaction between the cavity and the mirror is
\begin{equation}
\hat{\cal H}_{\rm int}=g_0/\sqrt{2} (\alpha_c^*+\delta\hat{c}^\dagger)(\alpha_c+\delta\hat{c}) (\alpha_m+\delta\hat{a}_m + \alpha_m^*+\delta\hat{a}_m^\dagger)
\end{equation}
which becomes
\begin{equation}
\hat{\cal H}_{\rm int} \simeq g_0/2 ( \alpha_c^*\delta\hat{c} + \alpha_c \delta\hat{c}^\dagger ) (\delta \hat{a}_m + \delta\hat{a}_m^\dagger) 
\label{Hint_noise}
\end{equation}
when we neglect the third order terms~\footnote{First order terms already cancel.}, e.g., $\sim \delta\hat{c}^\dagger \delta\hat{c}(\delta\hat{a}_m +\delta\hat{a}_m^\dagger)$, which corresponds to second order terms in the Langevin equations for noise operators~\cite{genes2008robust,vitaliPRL2007optomechanical}
\begin{eqnarray}
&&\delta \dot{\hat{x}}_m = \omega_m \delta \hat{p}_m \label{Langevin_noise_xm}
\\
&&\delta \dot{\hat{p}}_m = -\gamma_m \delta \hat{p}_m -\omega \delta\hat{x}_m + g_0 (\alpha_c^*\delta\hat{c} + \alpha_c\delta\hat{c}^\dagger) + g_m\epsilon_m(t) \nonumber
\\
\quad
\\
&&\delta\dot{\hat{c}} = (-\kappa+i\Delta)\delta\hat{c} + ig_0 \alpha_c \delta \hat{x}_m + g_c \hat{a}_{\rm in}(t), \label{Langevin_noise_c}
\end{eqnarray}
where $\hat{a}_{in}(t)$ and $\hat{\epsilon}_{in}(t)$ are input noise for the cavity and mechanical modes. $\Delta=\Delta_c-g_0 x_m$ appears in the steady state solution of Eqs.~(\ref{langevin_clas_xm})-(\ref{langevin_clas_alphac}). 
 
In the linearization procedure, both in Refs.~\cite{genes2008robust,vitaliPRL2007optomechanical} and Refs.~\cite{tarhan2013superluminal,agarwal2010electromagnetically}, second order terms are neglected in the Langevin equations. $g_c$ and $g_m$ are the coupling of the cavity mode and mechanical mode to the reservoirs and related to the damping parameters as $\gamma_c=\pi D(\omega_c) g_c^2$ and $\gamma_m=\pi \rho(\omega_m) g_m^2$ via input output relations~\cite{ScullyZubairyBook}~\footnote{Some researchers, e.g., Ref.~\cite{genes2008robust,vitaliPRL2007optomechanical,gardiner2004quantum}, use $g_c=\sqrt{\gamma_c}$ equivalently.}, where $D(\omega_c)$ and $\rho(\omega_m)$ are the optical and mechanical density of states. 

Writing Eq.~(\ref{Hint_noise}) in the form
\begin{equation}
\hat{\cal H}_{\rm int} = ( g^* \delta\hat{c} + g \delta\hat{c}^\dagger ) (\delta \hat{a}_m + \delta\hat{a}_m^\dagger),
\label{Hint_noiseG}
\end{equation}
with $g=\alpha_c g_0/2$, one can demonstrate that the laser pump is utilized for increasing the effective interaction between $\delta\hat{c}$ and $\delta\hat{a}_m$. 

Refs.~\cite{tarhan2013superluminal,agarwal2010electromagnetically} investigate the response of an optomechanical system to a weak probe pulse of frequency $\omega_p$ ($\Delta_p=\omega_p-\omega_{\rm L}$ in the rotating frame), around the steady-state, while the cavity is pumped by the coupler laser $\varepsilon_{\rm L}$. The weak probe pulse is included into the hamiltonian~(\ref{hamiltonian_full}) via an additional term ~\cite{agarwal2010electromagnetically,tarhan2013superluminal}
\begin{equation}
\hat{\cal H}_{\rm probe}=i\tilde{\varepsilon}_p(\hat{c}^\dagger e^{-i\Delta_p t}- \hat{c} e^{i\Delta_p t}).
\label{Hprobe}
\end{equation}  
Analogous to Langevin equations for noise operators, i.e., Eqs.~(\ref{Langevin_noise_xm})-(\ref{Langevin_noise_c}), Refs.~\cite{tarhan2013superluminal,agarwal2010electromagnetically} consider infinitesimally small probe fluctuation $\varepsilon_p=\tilde{\varepsilon}_p/g_c$ (dimensionless)
\begin{eqnarray}
\alpha_c(t) = \alpha_c + c_+ \varepsilon_p e^{-i\Delta_pt} + c_- \varepsilon_p^* e^{i\Delta_p t} \label{epsilonp_c}
\\
x_m(t)= x_m + x_+ \varepsilon_p e^{-i\Delta_p t} + x_- \varepsilon_p^* e^{i\Delta t}
\\
p_m(t)= p_m + p_+ \varepsilon_p e^{-i\Delta_p t} + p_- \varepsilon_p^* e^{i\Delta t} \label{epsilonp_pm}
\end{eqnarray}
over the steady-state amplitudes $\bar{\alpha}_c$, $\bar{x}_m$, $\bar{p}_m$ of the system driven by the strong coupler laser. The inclusion of the weak probe Eq.~(\ref{Hprobe}) changes merely the last equation in the equations of motion~(\ref{langevin_clas_xm})-(\ref{langevin_clas_alphac}) as
\begin{eqnarray}
\dot{\alpha}_c= -(\kappa+i\Delta_c)\alpha_c +i g_0 x_m \alpha_c + \varepsilon_{\rm L} + \tilde{\varepsilon}_p e^{-i\Delta_p t}. \label{langevin_clas_c_probe}
\end{eqnarray}

In the second-quantized treatment~\cite{genes2008robust,vitaliPRL2007optomechanical}, this corresponds to $\sim e^{\pm i \Delta_p t}$ fluctuations in $\delta\hat{c}$, $\delta\hat{x}_m$ and $\delta \hat{p}_m$ which can be introduced to the system, Eqs.~(\ref{Langevin_noise_xm})-(\ref{Langevin_noise_c}), via $\hat{a}_{in}(t)$ over the vacuum noise, i.e., $\hat{a}_{\rm in}(t)\to \hat{a}_{\rm in}(t) + \epsilon_p/g_c e^{-i\Delta_p t}$. 

Using Eqs.~(\ref{epsilonp_c})-(\ref{epsilonp_pm}) in the equations of motion~(\ref{langevin_clas_xm}),(\ref{langevin_clas_pm}) and (\ref{langevin_clas_c_probe}), one obtains the linear response of the system to the weak ($\varepsilon_p$) probe as~\cite{agarwal2010electromagnetically,tarhan2013superluminal}
\begin{equation}
\resizebox{.95\hsize}{!}{$
c_+=\frac{\left( [\kappa-i(\Delta+\Delta_p)](\Delta_p^2-\omega_m^2+i\gamma_m\Delta_p) -i\omega_m |g|^2 \right) }
{[(\kappa-i\Delta_p)^2+\Delta^2] (\Delta_p^2-\omega_m^2+i\gamma_m\Delta_p)+2\omega_m\Delta|g|^2} g_c \: .$
} 
\label{c+}
\end{equation}
Reflection $R(\omega)$ and transmission $T(\omega)$ coefficients can be determined from the input output relations,~\footnote{We kindly note that, in the calculation of Eqs.~(\ref{optomech_Rw}), (\ref{optomech_Tw}), one does not need the actual value of $g_c$. Because when the $g_c$, at the end of Eq.~(\ref{c+}), is included in Eqs.~(\ref{optomech_Rw}), (\ref{optomech_Tw}), $\gamma_c=\pi D(\omega_c) g_c^2$ appears.} 
\begin{eqnarray}
R(\omega) = 2\pi  D(\omega_c) g_c c_+ -1, \label{optomech_Rw}
\\
T(\omega)=2\pi D(\omega_c) g_c. \label{optomech_Tw}
\end{eqnarray}
For a single-sided cavity, i.e., when the right mirror is a perfect reflector attached to a mechanical oscillator, there is only reflection output. In this case, one merely needs to replace $\kappa=2\gamma_c$ with $\kappa=\gamma_c$, since the cavity couples to the reservoir from a single semitransparent mirror. $\gamma_c$ is the rate for the decay of the cavity field to outside (vacuum) from only one of the semitransparent mirrors.

\subsubsection{Nonanalyticities}

\begin{figure}
\begin{center}
\includegraphics[width=0.5\textwidth]{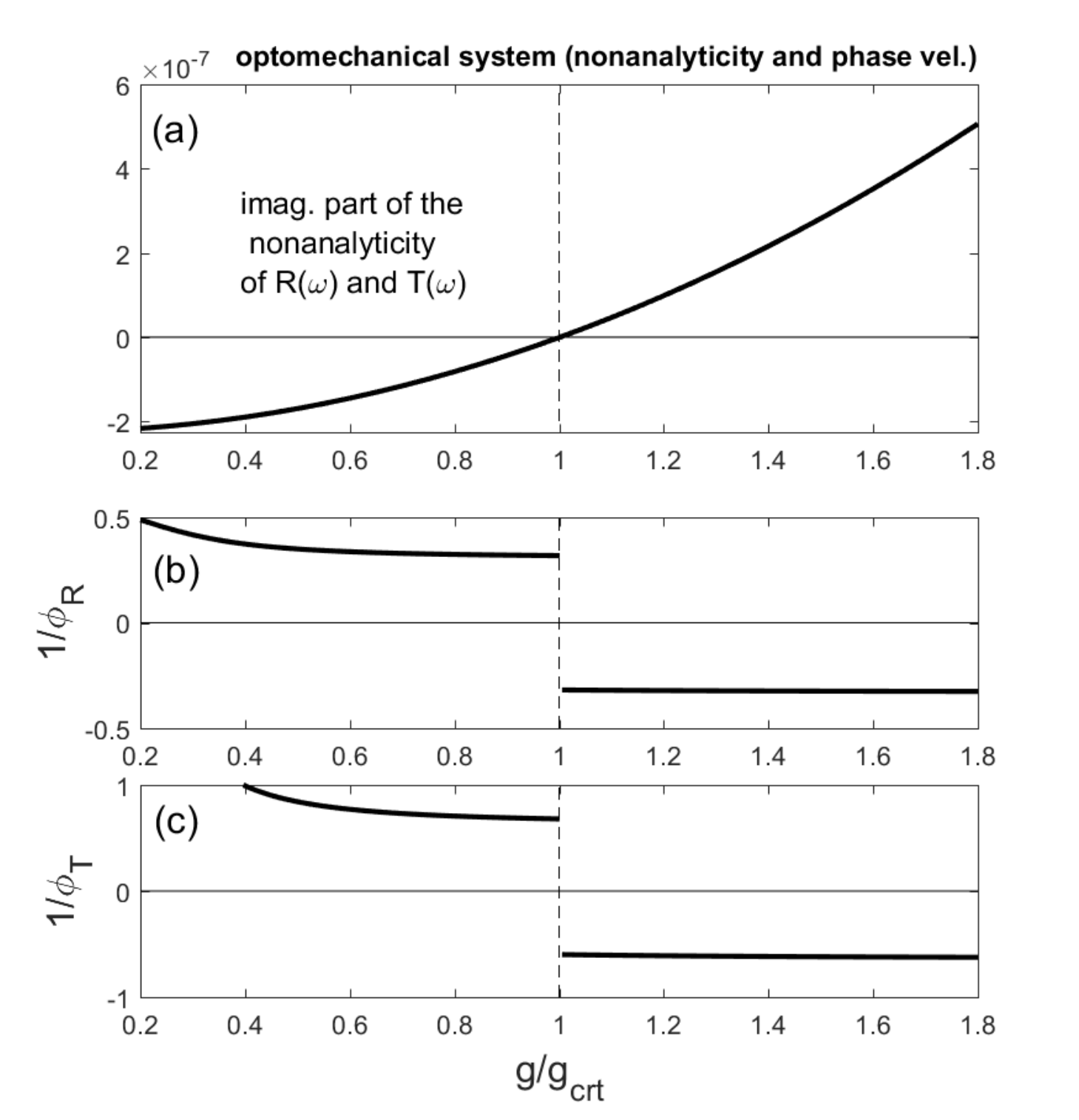}
\caption{ (a) Location of the nonanalyticities in the first-order response of an optomechanical system, depicted in Fig.~\ref{fig13}. Nonanalyticities of $R(\omega)$ move intp the upper half~(UH) of the complex frequency plane~(CFP), UH-CFP, after a critical cavity-mirror coupling $g>g_{\rm crt}$. Exactly at the same critical coupling phase velocity of the (b) reflected and (b) transmitted waves change sign from positive to negative abruptly. Thus, movement of the nonanalyticities in the first-order response is accompanied by negative phase (single frequency) velocities. This behavior is the same observed in Fig.~\ref{fig8} for an Otto configuration.
}
\label{fig14}
\end{center}
\end{figure}

In Fig.~\ref{fig14}a, we plot the locations of the nonnalyticities of $R(\omega)$. After exceeding a critical cavity-mirror coupling $g>g_{\rm crt}$, nonanalyticities of the ``first-order"  response function $R(\omega)$ move to the UH-CFP. Exactly at the same value $g_{\rm crt}$, negative phase velocities of reflected and transmitted waves accompany the violation of KKRs for that optomechanical system. This is the same behavior with Fig.~\ref{fig8} belonging to the Otto configuration.

\subsubsection{Nonanalyticities via  effective index method}

As a double-check, we also calculate the effective index~\cite{smith2002determination,yoo2019causal} for the optomechanical cavity described by the reflection $R(\omega)$ and transmission $T(\omega)$ functions. Effective dielectric function, again assuming a nonmagnetic medium, can be obtain from Eq.~(\ref{effective_eps}), given in Sec.~\ref{sec:effective_index}, as
\begin{equation}
\epsilon_{\rm eff}(\omega) = \frac{ [c_+(\omega)-2]^2 + c_+^2(\omega) e^{i2kL} }{ (1-e^{i2kL}) c_+^2(\omega) }.
\label{eps_eff_optome}
\end{equation}
$L$ is the cavity length and $k$ is the wavenumber.

A quick examination of the denominator of $\epsilon_{\rm eff}(\omega)$ in Eq.~(\ref{eps_eff_optome}) shows that $\epsilon_{\rm eff}(\omega)$ has poles (i) for $(1-e^{-i2kL})=0$ and (ii) for $c_+(\omega)=0$. Thus, effective index $n_{\rm eff}^2(\omega)=\epsilon_{\rm eff}(\omega)$ displays  nonanalyticities $c_+(\omega)=0$ which has nothing to do with the interference like origin, e.g., $(1-e^{-i2kL})=0$ whose solutions are already in the real-$\omega$ axis. In other words, $c_+(\omega)=0$, equivalently $T(\omega)=0$, do not depend on $L$~\footnote{\label{fn:Ldependence} Yes, $g_0$ depends on $L$. But it is not responsible for interference.}.

\begin{figure}
\begin{center}
\includegraphics[width=0.5\textwidth]{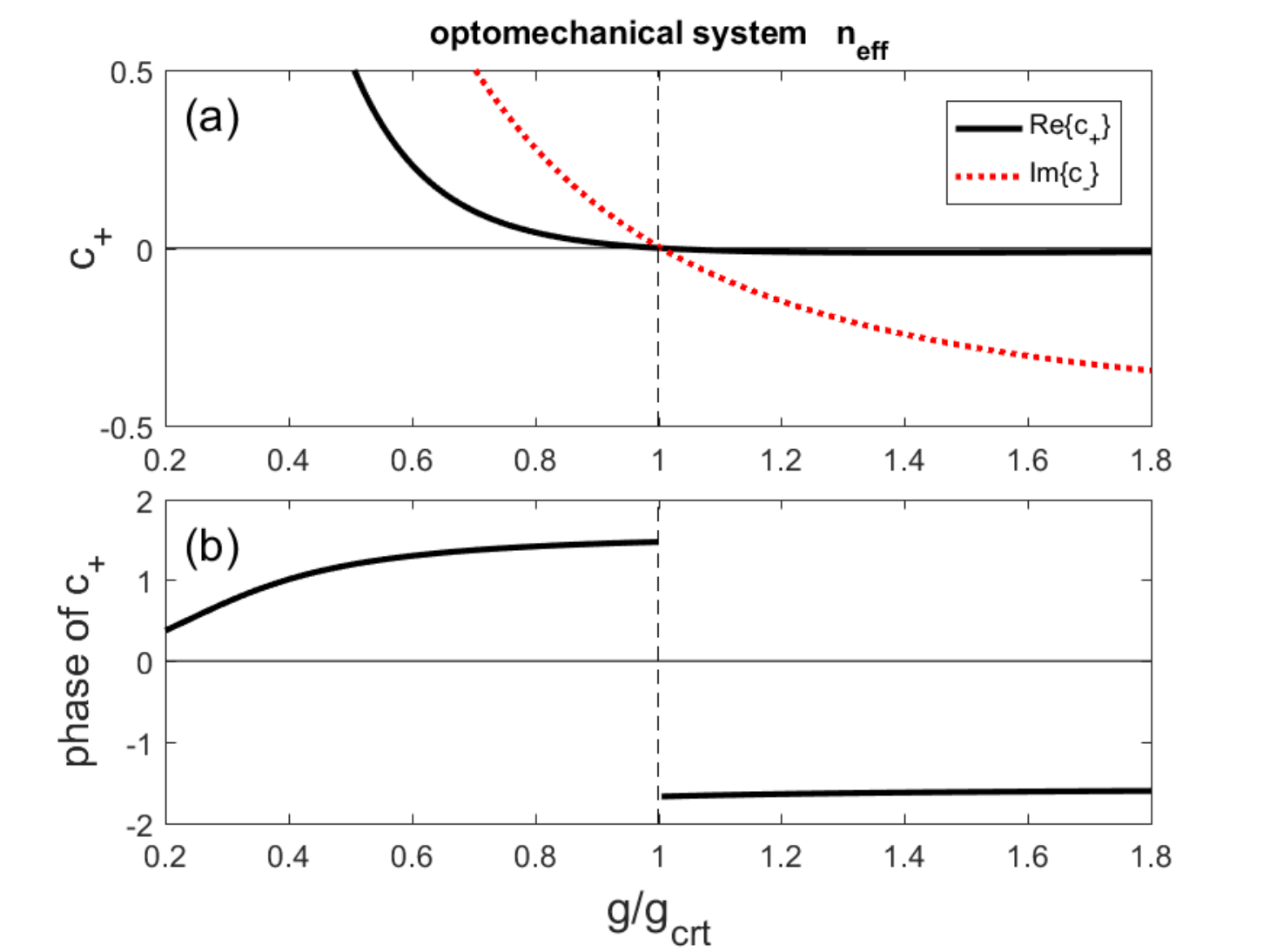}
\caption{ (a) Effective index $n_{\rm eff}(\omega) \propto 1/c_+(\omega)$ of an optomechanical system changes sign at the same critical coupling $g=g_{\rm crt}$  where nonanalyticities move into the UH-CFP in Fig.~\ref{fig14}a.  (b) The sign change in the phase of $n_{\rm eff}(\omega) \propto 1/c_+(\omega)$  indicates the presence of a negative index.
}
\label{fig15}
\end{center}
\end{figure}

Fig.~\ref{fig15}a demonstrates that effective index $n_{\rm eff}(\omega) \propto 1/c_+(\omega)$ changes sign at the critical coupling $g=g_{\rm crt}$ where first-order nonanalyticities of the system move to the UH-CFP in Fig.~\ref{fig14}a. Fig.~\ref{fig15}b further shows that phase of the effective index changes sign at the same critical coupling.

Although $\epsilon_{\rm eff}(\omega)$ in Eq.~(\ref{eps_eff_optome}) depends on $L$, actually, $L$ dependence can be completely removed from the system parameters if one considers a single-sided cavity. The effective index becomes
\begin{equation}
n_{\rm eff}(\omega) = \frac{1-R(\omega)}{1+R(\omega)} = \frac{2- 2\tilde{c}_+}{2\tilde{c}_+},
\end{equation}
where
\begin{equation}
\resizebox{.97\hsize}{!}{$
\tilde{c}_+(\omega)=  \frac{\left( [\kappa-i(\Delta+\Delta_p)](\Delta_p^2-\omega_m^2+i\gamma_m\Delta_p) -i\omega_m |g|^2 \right) }
{[(\kappa-i\Delta_p)^2+\Delta^2] (\Delta_p^2-\omega_m^2+i\gamma_m\Delta_p)+2\omega_m\Delta|g|^2} \gamma_c .
$}
\end{equation}
Here, again, $\tilde{c}_+(\omega)=\pi D(\omega_c) g_c c_{+}(\omega)=0$ determines the nonanalyticities (poles) and wave interference is not implemented in the system anymore~${}^{\ref{fn:Ldependence}}$.

While the origins of the violation of KKRs in Otto configuration~\cite{wang2016counterintuitive} and other systems~\cite{wang2002causal,beck1991group,stern2012transmission,wangPRA2007theoretical}, \cite{Zubairy2014counterintuitive} can be a jump-like~${}^{\ref{fn:jump}}$ behavior, the origin of the same phenomenon is the cavity-mirror coupling in optomechanics. This interaction is shown to induce a single-mode nonclassicality (e.g. quadrature squeezing) in the cavity mode above the same critical coupling, which is possible to survive ``violation of KKRs" from implementing the ``violation of causality"~\cite{tasgin2019entanglement}.

\begin{figure}
\begin{center}
\includegraphics[width=0.5\textwidth]{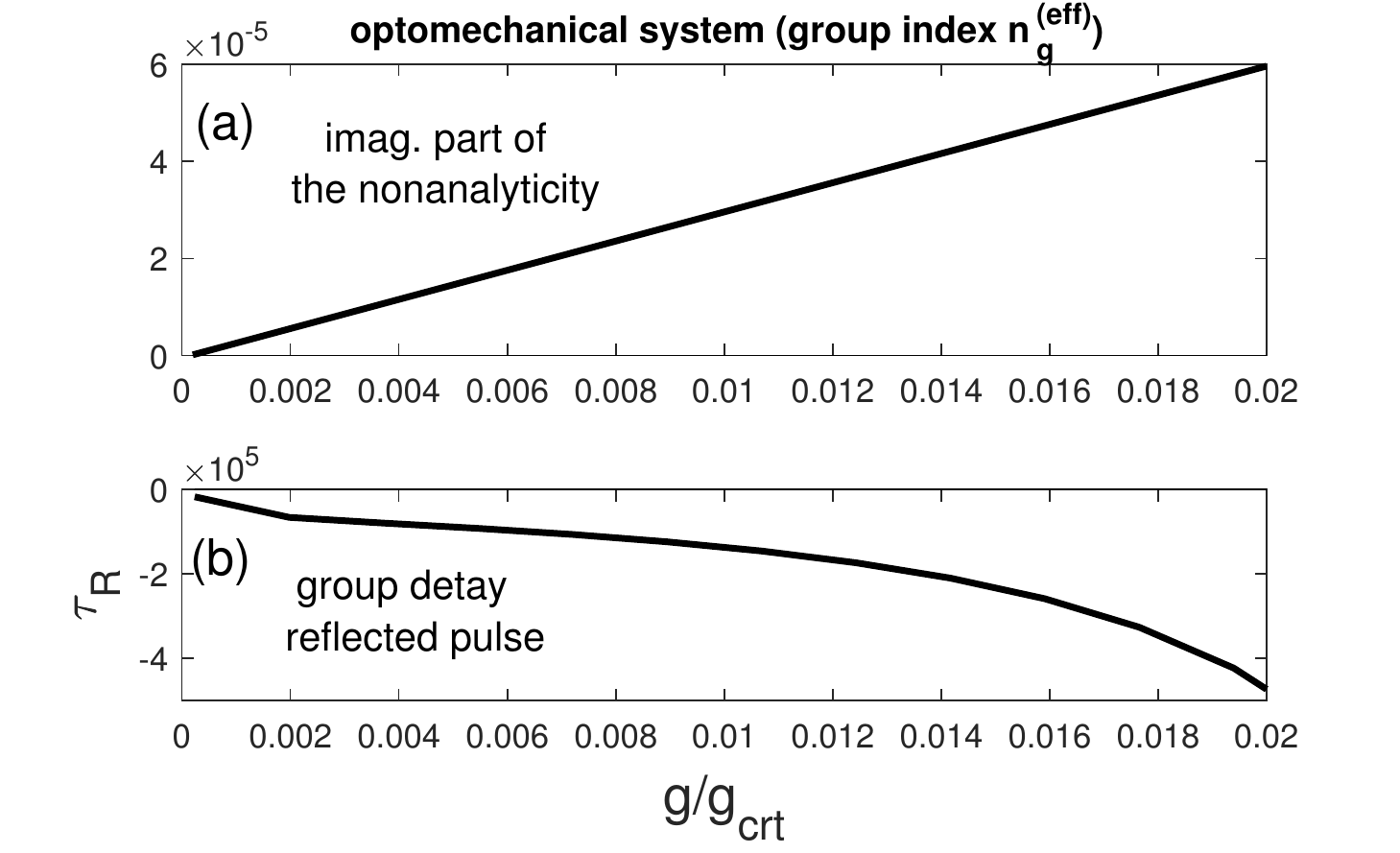}
\caption{(a) An optomechanical system also displays second-order nonanalyticities in the UH-CFP for the group response $n_{\rm eff}(\omega)$. (b) A negative group delay $\tau_R<0$, observed in Ref.~\cite{tarhan2013superluminal}, accompanies the presence of nonanalyticities in the UH-CFP.
}
\label{fig16}
\end{center}
\end{figure}

Above, we demonstrated the accompaniment of a negative phase velocity to the movement of the nonanalyticities to the UH-CFP. This is observed in the first-order response. Fig.~\ref{fig16} farther demonstrates that the negative group delay $\tau_R<0$, observed in Ref.~\cite{tarhan2013superluminal}, also accompanies the presence of the nonanalyticities of the group (second-order) response $\tau_R(\omega)$ in the UH-CFP.

\subsubsection{Gain medium?}

An optomechanical system actually is a gain medium. The coupler laser ($\varepsilon_{\rm L}e^{-i\omega_{\rm L}t}$) provides energy to the system which can provide gain to the weak probe pulse $\varepsilon_p e^{-i\Delta_p t}$. Then, one can naturally argue that movement of the nonanalyticities to the UH-CFP, in Fig.~\ref{fig14}, may also originate from this gain medium. That is, as it happens in an EIT medium~\cite{ScullyZubairyBook}.

As a counter-demonstration we kindly ask the audience to notice the following arguments. All the results presented in this subsection (i.e., Sec.~\ref{sec:optomechanics}), actually, is merely the solutions of the ``linearized hamiltonian" given by Eq.~(\ref{Hint_noiseG}). This is the same for Refs.~\cite{tarhan2013superluminal,agarwal2010electromagnetically,genes2008robust,vitaliPRL2007optomechanical}. After such a linearized treatment, the coupler laser pump ($\varepsilon_{\rm L} e^{-\omega_{\rm L}t}$) disappears from the effective hamiltonian which results the Langevin equations~(\ref{Langevin_noise_xm})-(\ref{Langevin_noise_c}) for the noise operators $\delta\hat{c}$ and $\delta\hat{a}_m$. The semi-classical approach~\cite{tarhan2013superluminal,agarwal2010electromagnetically}, Eqs.~(\ref{epsilonp_c})-(\ref{epsilonp_pm}), from which we obtain $c_+(\omega)$ in Eq.~(\ref{c+}), is also equivalent to inserting the fluctuations
\begin{eqnarray}
\delta c = c_+ \varepsilon_p e^{-i\Delta_pt}  + c_- \varepsilon_p e^{i\Delta_pt},
\\
\delta x_m = x_+ \varepsilon_p e^{-i\Delta_pt}  + x_- \varepsilon_p e^{i\Delta_pt},
\\
\delta p_m = p_+ \varepsilon_p e^{-i\Delta_pt}  + p_- \varepsilon_p e^{i\Delta_pt}  
\end{eqnarray} 
into the Langevin equations~(\ref{Langevin_noise_xm})-(\ref{Langevin_noise_c}) for the noise operators, as $\varepsilon_p << 1$ (or $\varepsilon_p\to\infty$).

That is, after the linearization, repeating ourselves that all the results do rely on, the laser pump intervene the system only by turning up the effective coupling $g=\alpha_c g_0$ in the hamiltonian~(\ref{Hint_noiseG}). We remark 3 points. (i) There is no gain in the Langevin equations~(\ref{Langevin_noise_xm})-(\ref{Langevin_noise_c}) governing $c_+(\omega)$. (ii) There is no physical restriction for $g$ in hamiltonian~(\ref{Hint_noiseG}) to be sufficiently large (i.e. $g>g_{\rm crt}$) without a cavity field enhancement. There is no such restrictions also for decay rates $\gamma_c$ and $\gamma_m$, i.e., no physical lower bound. (iii) $\hat{\cal H}_{\rm int}$ in Eq.~(\ref{Hint_noiseG}) is already a typical (physical) interaction hamiltonian for a two mode system. Therefore, (i)-(iii) demonstrate that the results presented in this subsection are also valid for a no-gain (for a sufficiently enough coupling or low damping) medium.

{\it In short}, in a two mode system interacting via hamiltonian~(\ref{Hint_noiseG}), a negative phase-velocity accompanies the movement of the nonanalyticities to the UH-CFP. We kindly note that our aim is the demonstration of such an accompaniment for different optical setups. And we observe that such an accompaniment appears both for a gain (active) and no-gain (passive) media as discussed in this subsection.

%%%%%%%%%%%%%%%%%%%%%%%%%%%%%%%%%%%%%%%%%%%%%%%%%%%%%%%%%%%%%%%%%%%%%%%%%%%%%%%%%%%%%%%%%%%%%%%%%%%%%%%%%%%%%%%%%%%%%%%%%%%%%%%%%%%%%%%%%%%%%%%%%%%%
%%%%%%%%%%%%%%%%%%%%%%%%%%%%%%%%%%%%%%%%%%%%%%%%%%%%%%%%%%%%%%%%%%%%%%%%%%%%%%%%%%%%%%%%%%%%%%%%%%%%%%%%%%%%%%%%%%%%%%%%%%%%%%%%%%%%%%%%%%%%%%%%%%%%
%%%%%%%%%%%%%%%%%%%%%%%%%%%%%%%%%%%%%%%%%%%%%%%%%%%%%%%%%%%%%%%%%%%%%%%%%%%%%%%%%%%%%%%%%%%%%%%%%%%%%%%%%%%%%%%%%%%%%%%%%%%%%%%%%%%%%%%%%%%%%%%%%%%%
\section{Summary and Conclusions} \label{sec:conclusion}

In summary, we present a systematic investigation of the nonanalyticities in the first-order and second-order responses of several optical setups. (i) We name the nonanalyticities as in the first-order if they appear in the refractive index or reflection/transmission functions. (ii) We name the nonanalyticities as in the second-order if they show up in the group (wave packet) behavior such as group index or group delay of a reflected beam.  Second-order response contains the derivatives of the first-order response functions with respect to $\omega$, e.g., $n_g(\omega)= dk/d\omega$ or $\tau_R(\omega)=d\phi_R/d\omega$. We explore: when such nonanalyticities move to the upper half~(UH) of the complex frequency plane~(CFP), UH-CFP, indicating a violation of Kramers-Kronig relations~(KKRs) ---not the causality itself~${}^{\ref{fn:causal}}$.

In short, we demonstrate that a negative velocity ($v<0$) accompanies the movement of the nonanayticities into the UH-CFP. This is observed both in the first-order and second-order responses. In the first-order response, we show that a jump of the phase velocity from positive to negative accompanies the movement of the nonanalyticities into the UH-CFP. The two transitions are shown to appear exactly at the same parameters for the Otto configuration and optomechanical system.

Regarding the second-order (group or wave packet) response, we show that a negative group delay (or velocity) accompanies the presence of nonanalyticities in the UH-CFP of the group response. 
Investigating  the nonanalyticities of the second-order response has particular importance. Because the pulse-center propagation, one measures in the experiments~\cite{ChuPRL1982SL,WangNature200SL,chiao1999tunneling,talukder2005measurement}, is governed by the group velocity~\cite{peatross2000average,nanda2009superluminal,talukder2014direct,kohmoto2005nonadherence}.

A crucial result we observe is: in a uniform Lorentzian dielectric, nonanalticities of the group index are not located in the UH-CFP for $v_g>c$. Nevertheless, the nonanalyticities are located in the UH-CFP when $v_g<0$. This result suggests the following understanding. $v_g>c$ (or $n_g<1$) is to be regarded analogous to $v>c$ (or $n<1$) for the phase velocity which is a common (not anomalous) phenomenon in optics. Because both for $v>c$ and $v_g>c$ nonanalyticities are in the UH-CFP, thus indicating the nonexistence of a true superluminal behavior (flow)~\cite{tasginPRA2012testing,talukder2014direct}. A true superluminal propagation would correspond to the violation of KKRs~(not the causality) due to the structure of the classical electromagnetism which is consistent with the special theory of relativity~\cite{Jackson_book,griffiths_book}. That is, a true superluminal flow would show itself as a flaw in electromagnetism.
 A negative index, $n_{\rm eff}<0$ or $n_g(\omega)<0$, however, is observed to be possible for possessing a true superluminal flow. Because the nonanalyticities move to the UH-CFP. 
 
 Therefore, while $v_g>c$ and $v_g<0$ are both referred as superluminal pulse-center propagation in the literature, we emphasize that, the two need to be differentiated from each other. $v_g<0$ is possible to be associated with a true superluminal flow ``in the group (second-order)" response while $v_g>c$ can be regarded analogous to phase velocity $v>c$. 
 
Our study is also related with the presence of negative-index materials without necessitating $\epsilon,\mu<0$~\cite{mackay2009negative,depine2004new,kinsler2008causality,stockman2007criterion}~${}^{\ref{fn:negative-index mater}}$. We believe that the phenomena (behaviors) we learn from the second-order response ($v_g$; the measured pulse-center propagation) can shed light also onto the first-order response.

The literature demonstrates the nonanalyticities appearing in interference-like devices~\cite{wang2002causal,beck1991group,stern2012transmission,wangPRA2007theoretical,Zubairy2014counterintuitive,wang2016counterintuitive} including the Otto configuration~\cite{wang2016counterintuitive}. In such systems, wave is possible to  vanish in some finite thickness spatial regions due to interference~${}^{\ref{fn:jump}}$. This can happen at certain wavelengths where reflected and transmitted waves cancel each other perfectly. Such a spatial gap may result anomalous tunneling-like (jump-like${}^{\ref{fn:jump}}$) behavior, where for instance light can behave as if tunneling two slabs of different thicknesses at equal time~\cite{hartman1962tunneling}. Hence, we expect that violation of KKRs in such interference-based setups could be regarded as appearing due to the assumption of instantaneous spreading of the wave functions to infinity~\cite{hegerfeldt1998instantaneous,hegerfeldtPRD1974remark,perezPRD1977localization,hegerfeldtPRD1980remarks} in wave mechanics.

In this work, we also demonstrate a phenomenon where movement of the nonanalyticities to the UH-CFP is not the interference. We show that an optomechanical system also violates the KKRs above a critical cavity-mirror coupling $g>g_{\rm crt}$ where, again, a negative phase-velocity introduces. A similar phenomenon is  demonstrated to appear also in a passive medium which mimics the effective interaction present in an optomechanical system.

Finally, we underline that we do not present a formal proof for the coexistence of negative phase/group velocity and the appearance of nonanalyticities in the UH-CFP. We rather demonstrate this accompaniment on several setups. Nevertheless, we believe that this demonstrations will stimulate further investigations on such a coincidence.

\bibliography{bibliography}
  
\end{document}